\documentclass[journal]{IEEEtran}
\usepackage{amsmath,graphicx}
\usepackage[linesnumbered,ruled]{algorithm2e}
\usepackage{multirow}
\usepackage{hhline}
\usepackage{cite}
\usepackage{pifont}
\usepackage{bm}
\usepackage{xcolor}
\newcommand{\tabincell}[2]{\begin{tabular}{@{}#1@{}}#2\end{tabular}}
\newcommand{\xmark}{\ding{55}}
\newcommand{\cmark}{\ding{51}}

\usepackage[pagebackref=false,colorlinks=true,bookmarks=false,linkcolor=red,citecolor=blue,urlcolor=red]{hyperref}

\ifCLASSINFOpdf
\else
\fi
\begin{document}
%
\title{Neural Architecture Search \\For LF-MMI Trained Time Delay Neural Networks}
%
%
%

\author{Shoukang Hu, Xurong Xie, Mingyu Cui$^*$, Jiajun Deng$^*$,  Shansong Liu, Jianwei Yu, Mengzhe Geng, \\Xunying Liu,~\IEEEmembership{Member,~IEEE}, Helen Meng,~\IEEEmembership{Fellow,~IEEE}
  \thanks{$^*$Two authors contributed equally. Shoukang Hu (e-mail: skhu@se.cuhk.edu.hk), Mingyu Cui, Jiajun Deng, Shansong Liu, Jianwei Yu, Mengzhe Geng, Xunying Liu (e-mail: xyliu@se.cuhk.edu.hk), Helen Meng (e-mail: hmmeng@se.cuhk.edu.hk) are with the Department of Systems Engineering and Engineering Management, The Chinese University of Hong Kong, Hong Kong SAR, China. Xurong Xie (e-mail: xurong@iscas.ac.cn) is with Institute of Software, Chinese Academy of Sciences, Beijing, China. (Corresponding author: Xunying Liu.)}
}

%
%


\markboth{IEEE/ACM Transactions on Audio, Speech and Language Processing}%
{Bare Demo of IEEEtran.cls for IEEE Journals}

%



\maketitle

\begin{abstract}
State-of-the-art automatic speech recognition (ASR) system development is data and computation intensive. The optimal design of deep neural networks (DNNs) for these systems often require expert knowledge and empirical evaluation. In this paper, a range of neural architecture search (NAS) techniques are used to automatically learn two types of hyper-parameters of factored time delay neural networks (TDNN-Fs): i) the left and right splicing context offsets; and ii) the dimensionality of the bottleneck linear projection at each hidden layer. These techniques include the differentiable neural architecture search (DARTS) method integrating architecture learning with lattice-free MMI training; Gumbel-Softmax and pipelined DARTS methods reducing the confusion over candidate architectures and improving the generalization of architecture selection; and Penalized DARTS incorporating resource constraints to balance the trade-off between performance and system complexity. Parameter sharing among TDNN-F architectures allows an efficient search over up to $7^{28}$ different systems. Statistically significant word error rate (WER) reductions of up to 1.2\% absolute and relative model size reduction of 31\% were obtained over a state-of-the-art 300-hour Switchboard corpus trained baseline LF-MMI TDNN-F system featuring speed perturbation, i-Vector and learning hidden unit contribution (LHUC) based speaker adaptation as well as RNNLM rescoring. Performance contrasts on the same task against recent end-to-end systems reported in the literature suggest the best NAS auto-configured system achieves state-of-the-art WERs of 9.9\% and 11.1\% on the NIST Hub5’ 00 and Rt03s test sets respectively with up to 96\% model size reduction. Further analysis using Bayesian learning shows that the proposed NAS approaches can effectively minimize the structural redundancy in the TDNN-F systems and reduce their model parameter uncertainty. Consistent performance improvements were also obtained on a UASpeech dysarthric speech recognition task. Our code is available at \url{https://github.com/skhu101/TDNN-F_NAS}.

\end{abstract}


%
\IEEEpeerreviewmaketitle

\section{Introduction}
\IEEEPARstart{S}{tate-of-the-art} automatic speech recognition (ASR) system are becoming increasingly complex. Deep learning techniques play a key role in these systems~\cite{waibel1989consonant, kingsbury2009lattice, vesely2013sequence, su2013error,  peddinti2015time, povey2016purely, povey2018semi} and have in recent years evolved into a large set of advanced deep neural network (DNN) models. These include the traditional hybrid HMM-DNN architecture~\cite{povey2018semi,abdel2012applying, abdel2013exploring,medsker2001recurrent,graves2013speech,amberkar2018speech} featuring convolutional neural networks (CNNs)~\cite{abdel2012applying, abdel2013exploring}, time delay neural networks (TDNNs)~\cite{waibel1989consonant, peddinti2015time, povey2018semi, povey2016purely} or recurrent neural networks (RNNs)~\cite{medsker2001recurrent,graves2013speech,amberkar2018speech} and their long short-term memory variants~\cite{hochreiter1997long, sak2014long}; and the recently emerging all neural end-to-end (E2E) modelling paradigm represented by listen, attend and spell (LAS)~\cite{chan2016listen}, connectionist temporal classification (CTC)~\cite{graves2006connectionist}, RNN transducers (RNN-T)~\cite{graves2012sequence} and neural transformer models~\cite{vaswani2017attention,DBLP:conf/naacl/DevlinCLT19,dong2018speech,gulati2020conformer}.

The development of these systems is data and computation intensive. The optimal design of neural architectures in these systems often requires a large set of hyper-parameters encoding varying structural configurations to be set, for example, the hidden layer dimensionality and connectivity between different layers. To date these are largely determined based on expert knowledge or empirical choice. As explicitly training and evaluating the performance of all possible neural structural configurations is highly expensive, the need of deriving suitable automated neural network architecture learning techniques~\cite{baluja1994reducing, kirschning1995parallel} for speech recognition systems becomes particularly salient. 

To this end, neural architecture search (NAS) approaches~\cite{elsken2019neural} have gained increasingly interest in recent years in both the computer vision~\cite{zoph2016neural, pham2018efficient, real2017large, liu2018hierarchical, elsken2018efficient, DBLP:conf/nips/KandasamyNSPX18, liu2018darts, xie2018snas, wu2019fbnet, cai2018proxylessnas, dong2019searching, hu2020dsnas, DBLP:conf/iclr/ChenWCTH21, xie2020understanding, real2019regularized, tan2019mixconv, Liu_2019_CVPR, Nekrasov_2019_CVPR,chen2019detnas,tan2019efficientnet} and speech~\cite{veniat2019stochastic,mo2020neural, ding2020autospeech,li2019neural,moriya2018evolution,kim2020evolved,chen2020darts,he2021learned,zheng2021efficient,liu2021improved,shi2021efficient,mehrotra2021bench,hu2021neural, deng21d_interspeech} communities. The key objectives of NAS methods can be formulated as three fold. First, to allow the best system to be selected, it is crucial for NAS methods to produce an accurate performance rank ordering over different candidate neural architectures. Second, when designing practical systems operating on a given ASR accuracy performance target, preference should be given to neural architectures with fewer parameters in order to reduce the parameter uncertainty and minimize the risk of overfitting to limited training data. Furthermore, to ensure the NAS algorithms’ scalability and efficiency on large data sets, a compact search space containing all candidate neural architectures of interest and exploiting the structural commonalities among them needs to be defined. 

Earlier forms of NAS techniques were based on neural evolution~\cite{angeline1994evolutionary,stanley2002evolving, floreano2008neuroevolution,stanley2009hypercube,jozefowicz2015empirical}, where genetic algorithms were employed during mutation and crossover rounds by randomly selecting architecture choices. Performance and efficiency of these evolution based NAS methods heavily depends on the precise choice over parent model structures, mutation population groups and off-springs. For example, a pairwise competition based tournament selection approach~\cite{DBLP:conf/foga/GoldbergD90} was used to sample parents in~\cite{real2017large, liu2018hierarchical}, while a Pareto optimality based multi-objective approach was used in~\cite{elsken2018efficient} to select neural architectures and adjust the trade-off between the predicted performance and number of parameters. NAS methods based on Bayesian Optimization (BO)~\cite{shahriari2015taking} were also explored to predict the performance rank ordering. Performance ranking prediction over candidate architectures was learned in non-parametric fashion using Gaussian Process (GP) models~\cite{rasmussen2006gaussian} in~\cite{swersky2014raiders,DBLP:conf/nips/KandasamyNSPX18}, where the similarity metric among candidate architectures was efficiently computed using kernel functions. 

Reinforcement learning (RL) based NAS approaches~\cite{zoph2016neural, pham2018efficient} have also been investigated. In these methods, the search space is formulated as a discrete state space and the generation of neural architectures is regarded as the agent’s actions in response to the reward based on the performance of the sampled models. Explicit system training and evaluation of candidate architectures are required in the above existing NAS techniques. In addition, as the architecture hyper-parameters and actual DNN parameters are separately learned, for example, within the RL controller and candidate neural network systems, a tighter integration of both is preferred.

Alternatively, differentiable neural architectural search (DARTS) techniques~\cite{liu2018darts, xie2018snas, wu2019fbnet, cai2018proxylessnas, dong2019searching, hu2020dsnas, DBLP:conf/iclr/ChenWCTH21, xie2020understanding} can be used. Neural architecture search is efficiently performed over an over-parameterized super-network model that contains all possible candidate architectures to be considered. Within such a super-network, the underlying NAS problem is transformed into the estimation of the weight parameters assigned to each candidate neural architecture. After the super-network model containing both architecture weights and normal DNN parameters has been trained to convergence, the optimal architecture is obtained by pruning lower weighted paths. One prominent advantage of DARTS based NAS methods is that they allow both the architecture selection and candidate DNN parameters to be consistently optimized within the same super-network model. 

In this paper, a range of DARTS based NAS techniques are used to automatically learn two architecture hyper-parameters that heavily affect the performance and model complexity of state-of-the-art lattice-free Maximum Mutual Information (LF-MMI) trained factored time delay neural network (TDNN-F)~\cite{waibel1989consonant, peddinti2015time, povey2018semi, povey2016purely} acoustic models: i) the left and right splicing context offsets; and ii) the dimensionality of the bottleneck linear projection at each hidden layer. These include the standard DARTS method that fully integrates the estimation of architecture weights and TDNN-F parameters in LF-MMI training; the Gumbel-Softmax DARTS technique producing approximately one-hot architecture distributions to reduce the confusion during model search; the pipelined DARTS method that circumvents the overfitting of architecture weights using validation data; and the penalized DARTS approach that further incorporates a resource penalty to flexibly adjust the trade-off between performance and system complexity. Parameter sharing among candidate architectures allows efficient search over a large number (up to $7^{28}$) of different TDNN-F systems to be performed. 

Experiments conducted on a state-of-the-art 300-hour Switchboard corpus trained baseline LF-MMI TDNN-F system featuring speed perturbation, i-Vector and learning hidden unit contribution (LHUC)~\cite{swietojanski2016learning} based speaker adaptation as well as RNNLM rescoring suggest the NAS configured TDNN-F models consistently outperform the baseline systems using manually designed configurations or random architecture search. Significant absolute word error rate (WER) reductions up to 1.2\% and model size reduction of 31\% relative were obtained. Performance contrasts on the same task against most recent hybrid and end-to-end attention and transformer based systems reported in the literature~\cite{kitza2019cumulative,karita2019comparative,park2019specaugment,tuske2020single} suggest our best NAS auto-configured system achieve state-of-the-art WERs of 9.9\% and 11.1\% on the NIST Hub5’00 and Rt03s test sets respectively with up to 96\% model size reduction. Further analysis using Bayesian learning shows the proposed NAS approaches can effectively minimize the structural redundancy in the TDNN-F systems and reduce their model parameter uncertainty. Consistent performance improvements were also obtained on a UASpeech~\cite{kim2008dysarthric} dysarthric speech recognition task.

The main contributions of this paper are summarized below: 
\begin{enumerate}
    \item This paper presents the first use of DARTS based NAS techniques to automatically learn architecture hyper-parameters that directly affect the performance and model complexity of state-of-the-art LF-MMI trained TDNN-F acoustic models. In contrast, previous NAS researches conducted on similar systems either used a) evolutionary algorithms requiring expert setting of initial genes and long evaluation time for each individual candidate architecture~\cite{moriya2018evolution} (up to 4 days even with a manual early-stopping mechanism) while in our NAS approaches the entire architecture search is performed over all possible $7^{28}$ candidate systems and model training cycle is limited to approximately 6.6 GPU days; or b) architecture sampling based straight-through gradient approach~\cite{zheng2021efficient} on a TDNN-CTC end-to-end system producing much higher WERs (12.6\% and 23.2\%) on the swbd and callhm subsets of Hub5’ 00 test set than our NAS auto-configured TDNN-F systems on the same data (6.9\% and 13.0\%) presented in this paper.
    \item To facilitate efficient search over a very large number of TDNN-F systems, this paper presents the first use of a flexible model parameter sharing scheme that is tailor-designed for specific hyper-parameters contained in TDNN-F systems, to the best of our knowledge. The generic nature of the proposed NAS methods accompanying parameter sharing technique also allows them to be employed to improve the efficiency and scalability of similar neural architecture design issues encountered during system development for end-to-end approaches including transformers~\cite{liu2021improved,shi2021efficient}.
    \item This paper presents the best published NAS based LF-MMI TDNN-F system performance reported in the literature on the 300-hour Switchboard task to the best of our knowledge. Performance contrasts on the same task against most recent hybrid and end-to-end attention and transformer based systems reported in the literature~\cite{kitza2019cumulative,karita2019comparative,park2019specaugment,tuske2020single} suggest our best NAS auto-configured system achieves state-of-the-art WERs of 9.9\% and 11.1\% on the NIST Hub5’00 and Rt03s test sets respectively with model size reduction of up to 96\% relative.
    \item This paper further presents the earliest work on analysing the efficacy of NAS approaches when being used to minimize the structural redundancy in the TDNN-F systems and reduce their model parameter uncertainty when given limited training data. In contrast, only speech recognition accuracy performance and model size reduction are investigated in previous researches~\cite{moriya2018evolution,kim2020evolved,chen2020darts,he2021learned,zheng2021efficient,liu2021improved,shi2021efficient,mehrotra2021bench,tuske2020single}.
\end{enumerate}

\subsection{Related works}
With the rapid development of NAS techniques in the machine learning and computer vision communities, there has been also increasingly interest in applying these to speech recognition systems.
\begin{itemize}
    \item Evolutionary algorithms were used in~\cite{moriya2018evolution} to automatically learn a series of architecture hyper-parameters in time delay neural networks (TDNNs)~\cite{waibel1989consonant, peddinti2015time}, and achieved 0.9\% absolute WER reduction and 36\% relative model size reduction over the Kaldi recipe trained baseline system on a spontaneous Japanese speech recognition task~\cite{furui2000japanese}.
    \item Genetic algorithms were also used in~\cite{kim2020evolved} to automatically configure transformer model architectures. The NAS configured transformer achieves 12.9\% WER on the Wall Street Journal (WSJ) task, outperforming the baseline transformer system by 0.6\% absolute WER reduction.
    \item DARTS based NAS method have been investigated to automatically configure hyper-parameters for CNNs~\cite{chen2020darts,he2021learned}, ST-NAS~\cite{zheng2021efficient} and transformer models~\cite{liu2021improved,shi2021efficient} respectively. 
    In particular, the concurrent work ST-NAS~\cite{zheng2021efficient} applied an architecture sampling based straight-through gradient approach on a TDNN-CTC end-to-end system. WERs of 12.6\% and 23.2\% were reported on the swbd and callhm subset of the Hub5’ 00 test set on the Switchboard conversational telephone speech recognition task.
    The ST-NAS configured system outperformed the baseline system by 2.0\% and 2.3\% absolute WER reduction.
    \item  Performance of various NAS techniques were also evaluated in~\cite{mehrotra2021bench} on the TIMIT task for CNN models.
\end{itemize}

The rest of this paper is organized as follows. In section 2, a set of differentiable neural architecture search (DARTS) techniques are first introduced. These include the standard DARTS, Gumbel-Softmax DARTS, pipelined DARTS and penalized DARTS methods. Time delay neural networks are reviewed in Section 3. The search spaces for automatically learning the settings of context offsets and bottleneck projection dimensionality choices in TDNN-F models and the accompanying efficient parameter sharing scheme are discussed in Section 4. Section 5 shows the experiments and results. Finally, the conclusions are drawn in Section 6.

\section{Neural Architecture Search}
\label{NAS_methods}
In this section, we introduce several forms of differentiable neural architecture search (DARTS) methods considered in this paper, including the standard DARTS, Gumbel-Softmax DARTS, pipelined DARTS and penalized DARTS methods.

With no loss of generality, the general form of DARTS architecture selection methods~\cite{liu2018darts, xie2018snas, cai2018proxylessnas, hu2020dsnas} are introduced as follows. For example, the $l$-th neural network hidden layer output $\mathbf{h}^l$ can be computed as a linear combination between the architecture weights $\lambda_i^l$ and candidate architecture choices $\phi_i^l(\cdot)$ in the DARTS super-network:
\begin{equation}
\label{eq:darts_fw}
\setlength{\abovedisplayskip}{4pt}
\setlength{\belowdisplayskip}{4pt}
\begin{aligned}
\mathbf{h}^l=&\sum_{i=0}^{N^l-1}\lambda_i^l\phi_i^l(\mathbf{W}_i^{l}\mathbf{h}^{l-1})
\end{aligned}
\end{equation}
where $\lambda_i^l$ is the architecture weight for the $i$-th candidate choice in the $l$-th layer, $N^l$ is the total number of choices in this layer. The precise form of neural architectures being considered at this layer is determined by the linear transformation parameter $\mathbf{W}_i^{l}$ and activation function $\phi_i^l(\cdot)$ used by each candidate system. For example, when selecting the TDNN-F hidden layer context offsets, the linear transformation is a binary-valued matrix for each candidate architecture, and $\phi_i^l(\cdot)$ is represented by an identity matrix. When selecting the dimensionality of the bottleneck linear projection at each hidden layer, the linear transformation $\mathbf{W}_i^{l}=\widetilde{\mathbf{W}}_i^l\widehat{\mathbf{W}}_i^{l^T}$ is a decomposed matrix, while $\phi_i^l(\cdot)$ is also an identity matrix.

\vspace{-3mm}
\subsection{Softmax DARTS}
In the conventional DARTS~\cite{liu2018darts} system, a Softmax function is used to model the architecture selection weights $\lambda_i^l$:
\begin{equation}
\label{eq:softmax_fw}
\setlength{\abovedisplayskip}{4pt}
\setlength{\belowdisplayskip}{4pt}
\begin{aligned}
\lambda_i^l=\frac{\exp(\log\alpha_i^l)}{\sum_{j=0}^{N^l-1}\exp(\log\alpha_j^l)}
\end{aligned}
\end{equation}
where $\log\alpha_i^l$ is the architecture dependent parameter determining their contribution during NAS search.

When using the standard back-propagation algorithm to update the architecture weights parameter $\lambda_i^l$, the loss function (including LF-MMI criterion~\cite{povey2016purely} considered in this paper) gradient against the $\log\alpha_k^l$ is computed as below.
\begin{equation}
\label{eq:softmax_bp}
\begin{aligned}
\frac{\partial\mathcal L}{\partial\log\alpha_k^l}=\frac{\partial\mathcal{L}}{\partial\mathbf{h}^l}\sum_{i=0}^{N^l-1}\left(1_{i=k}\lambda_i^l-\lambda_i^{l}\lambda_k^{l}\right)\phi_i^l(\mathbf{W}_i^{l}\mathbf{h}^{l-1})
\end{aligned}
\end{equation}
where $1_{i=k}$ is the indicator function. When the DARTS super-network containing both architecture weights and normal DNN parameters is trained to convergence, the optimal architecture can be obtained by pruning lower weighted architectures that are considered less important. However, when similar architecture weights are obtained using a flattened Softmax function, the confusion over different candidate systems increases and search errors may occur.

\begin{figure}[t]
    \vspace{-3mm}
    \centering
    \includegraphics[width=3.5in]{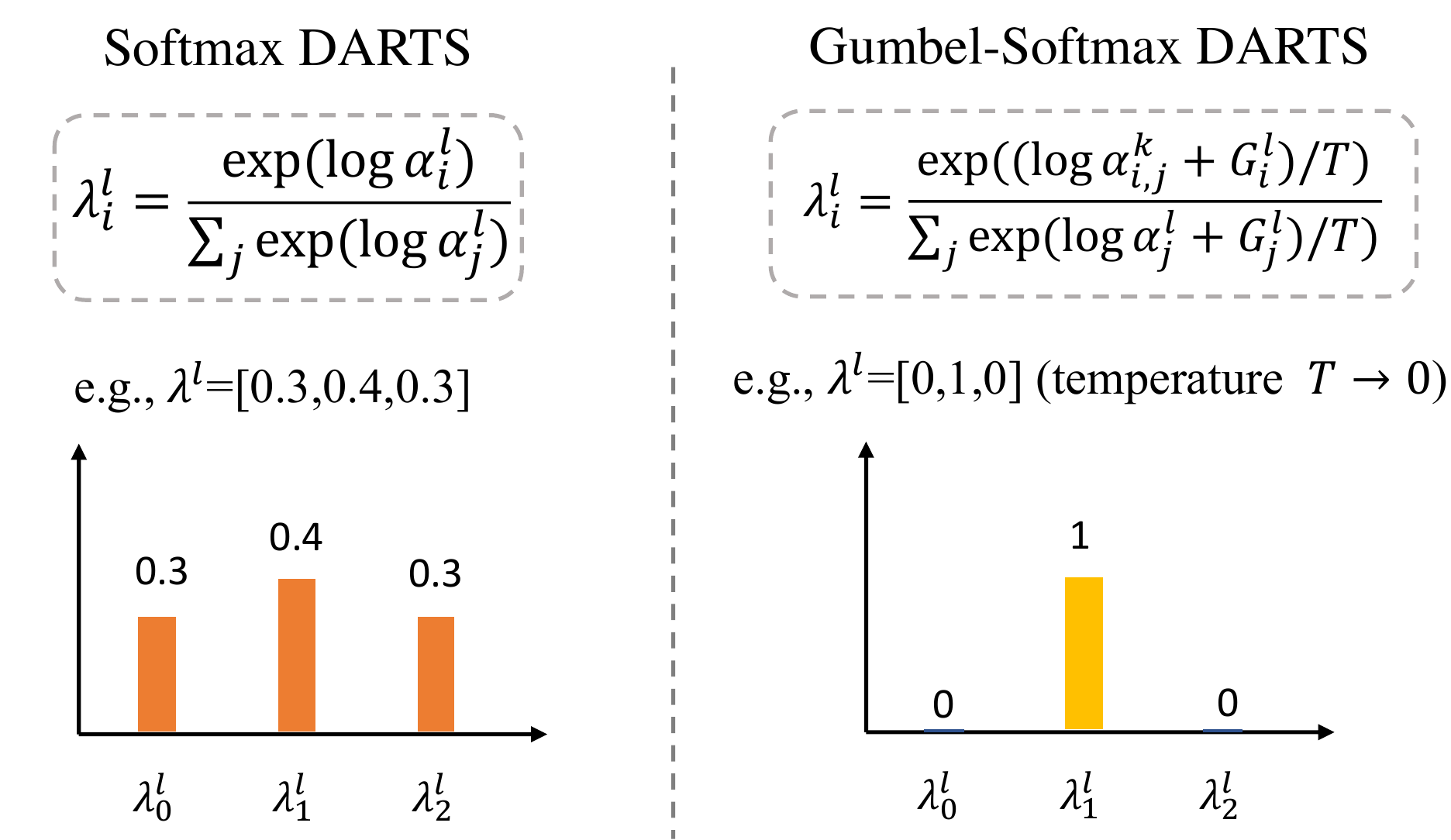}
    \vspace*{-6.5mm}
    \caption{Example architecture weights learned using Softmax DARTS (left) and Gumbel-Softmax DARTS (right) respectively.}
    \label{fig: softmax_gumbel_softmax}
    \vspace{-5mm}
\end{figure}

\vspace{-3mm}
\subsection{Gumbel-Softmax DARTS}
In order to address the above issue, a Gumbel-Softmax distribution~\cite{maddison2016concrete, xie2018snas, wu2019fbnet, dong2019searching} is used to sharpen the architecture weights to produce approximately a one-hot vector. This allows the confusion between different architectures to be minimised. The architecture weights are computed as, 
\begin{equation}
\label{eq:gumbel_reparam}
\begin{aligned}
\lambda_{i}^{l}=\frac{\exp((\log\alpha_i^l+G_{i}^l)/T)}{\sum_{j=0}^{N^l-1}\exp((\log\alpha_j^l+G_{j}^l)/T)}
\end{aligned}
\end{equation}
where $G_{i}^l=-\log(-\log(U_{i}^l))$ is the Gumbel variable, and $U_{i}^l$ is a uniform random variable. When the temperature parameter $T$ approaches 0, it has been shown that the Gumbel-Softmax distribution is close to a categorical distribution~\cite{maddison2016concrete}. An example contrast between the Softmax DARTS and Gumbel-Softmax DARTS methods learned neural architecture weights is shown in Fig.~\ref{fig: softmax_gumbel_softmax}.

Different samples of the uniform random variable $U_i^l$ lead to different values of $\lambda_i^l$ in Eq.~\ref{eq:gumbel_reparam}. The loss function gradient w.r.t $\log\alpha_k^l$ is computed as an average over $J$ samples of the architecture weights,
\begin{equation}
\label{eq:gumbel_bp}
\setlength{\abovedisplayskip}{4pt}
\setlength{\belowdisplayskip}{4pt}
\begin{aligned}
\frac{\partial\mathcal{L}}{\partial\log\alpha_k^l}=\frac{1}{J}\sum_{j=0}^{J}\frac{\partial\mathcal L}{\partial\mathbf{h}^{l,j}}\sum_{i=0}^{N^l-1}\frac{1_{i=k}\lambda_{i}^{l,j}-\lambda_{i}^{l,j}\lambda_{k}^{l,j}}{T}\phi_i^l(\mathbf{W}_i^{l}\mathbf{h}^{l-1,j})
\end{aligned}
\end{equation}
where $\bm{\lambda}^{l,j}$ is the $j$-th sample weights vector drawn from the Gumbel-Softmax distribution in the $l$-th layer, $\mathbf{h}^{l,j}$ is the output of $l$-th layer by using the $j$-th sample $\bm{\lambda}^{l,j}$. We assume the Gumbel-Softmax variables $\bm{\lambda}^{l}$ at different layers are mutually independent.

\vspace{-3mm}
\subsection{Pipelined DARTS}
As both architecture weights and normal DNN parameters are learned at the same time in Softmax and Gumbel-Softmax DARTS systems, the search algorithms may prematurely select sub-optimal architectures at an early stage. Inspired by~\cite{guo2020single}, we decouple the update of normal DNN parameters and architecture weights into two separate stages performed in sequence. This leads to the pipelined DARTS approach. In order to prevent overfitting to the training data, a separate held-out data set taken out of the original training data is used. In Pipelined DARTS systems, the normal DNN parameters are updated to convergence on the training data first (not containing the separate held-out data), while randomly sampled one-hot architecture weights drawn from a uniform distribution are used. In the following stage, we fix the normal DNN parameters estimated in the first stage in the super-network and update the architecture weights using the held-out data for both Softmax DARTS and Gumbel-Softmax DARTS. This 
produces the Pipelined Softmax DARTS (PipeSoftmax) and Pipelined Gumbel-softmax DARTS (PipeGumbel) systems.

\vspace{-3mm}
\subsection{Penalized DARTS}
\label{sec: penalized_darts}
In order to flexibly adjust the trade-off between system performance and complexity, a penalized loss function incorporating the underlined neural network size is used, given as follows:
\begin{equation}
\label{eq:penalized_darts}
\setlength{\abovedisplayskip}{4pt}
\setlength{\belowdisplayskip}{4pt}
\begin{aligned}
\mathcal{L} = \mathcal{L}_{LF-MMI} + \eta \sum_{l,i}\lambda_{i}^{l}C_{i}^l
\end{aligned}
\end{equation}
where $\mathcal{L}_{LF-MMI}$ is the lattice-free MMI criterion considered in this paper. $C_{i}^l$ is the term related with the model complexity of the $i$-th candidate considered at the $l$-th layer, which can be expressed in different forms, for example, the number of model parameters, floating-point operations (FLOPs) or the latency computed given the specified hardware. Unless otherwise stated in this paper, we treat $C_{i}^l$ as the number of parameters of the $i$-th candidate considered at the $l$-th layer. $\eta$ is the penalty scaling factor empirically set for different tasks. 

\section{Time Delay Neural Network}
Time delay neural networks (TDNNs)~\cite{waibel1989phoneme, waibel1989consonant, peddinti2015time, povey2018semi, povey2016purely, hadian2018end} based hybrid HMM-DNN acoustic models in recent years defined state-of-the-art speech recognition performance over a wide range of tasks, due to their strong power in modelling long range temporal dependencies in speech. In particular, the recently proposed factored TDNN systems~\cite{povey2018semi} featuring lattice-free MMI sequence discriminative training~\cite{povey2016purely} remain highly competitive against all neural end-to-end approaches to date~\cite{hadian2018end, luscher2019rwth, karita2019comparative,park2019specaugment,tuske2020single, zhou2020rwth, li2020comparison}.
\begin{figure}[h]
    \vspace{-3mm}
    \centering
    \centerline{\includegraphics[width=3.5in]{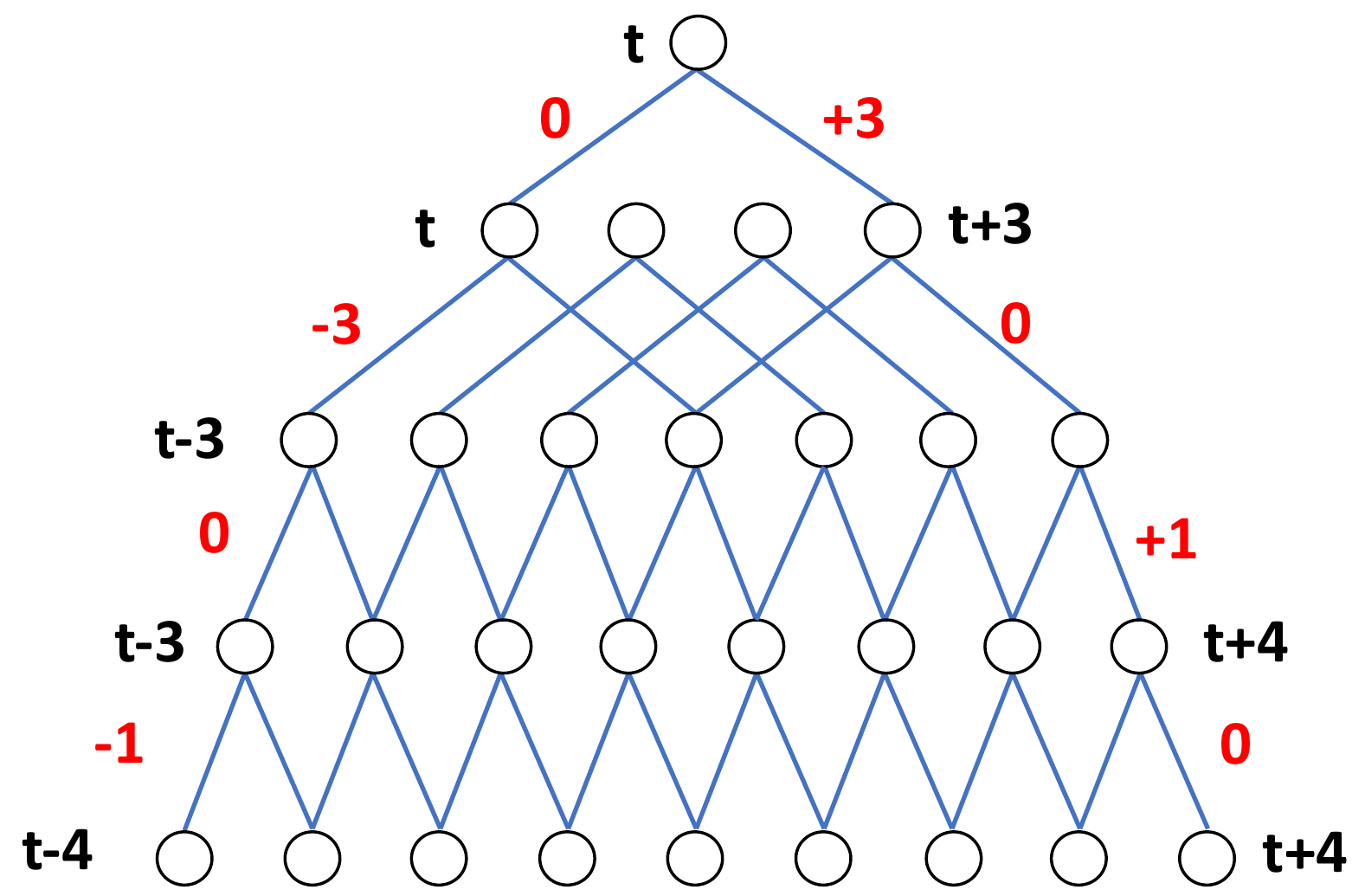}}
    \vspace*{-3mm}
    \caption{An example TDNN architecture with splicing context offsets \{-1,0\} \{0,1\} \{-3,0\} \{0,3\} in turn from the bottom to the top layer.}
    \label{fig: TDNN}
    \vspace{-3mm}
\end{figure}

TDNNs can be considered as a special form of one-dimensional convolutional neural networks (CNNs)~\cite{lecun1995convolutional} when parameters are tied across different time steps. An example TDNN model is shown in Fig.~\ref{fig: TDNN}. The bottom layers of TDNNs are designed to learn a narrower temporal context span, while the higher layers to learn wider, longer range temporal contexts. One important type of hyper-parameters in TDNN models controlling its temporal modelling ability is the left and right splicing context offsets. These alter the temporal context ranges learned in each TDNN hidden layer. The splicing context offsets used in the example of Fig.~\ref{fig: TDNN} are \{-1,0\} \{0,1\} \{-3,0\} \{0,3\} from the bottom to the top layer. To further reduce the risk of overfitting to limited training data and the number of parameters, a factored TDNN (TDNN-F) model structure was proposed in~\cite{povey2018semi}, which compresses the weight matrix by using semi-orthogonal matrix decomposition. In this TDNN-F model, the hidden layer specific bottleneck projection dimensionality settings present another group of hyper-parameter that needs to be determined.


If we assume 7 possible context offset choices to the left and another 7 context offset choices to the right to be learned at each of the 14 TDNN-F layers, there are up to $7^{28}$ context offset choices being considered in the search space. When determining the number of bottleneck dimensions, for example, out of a total of 8 possible settings, at each of the 14 hidden TDNN-F layers, again a very large number of possible system configurations of $8^{14}$ need to be considered during NAS. In this paper, we adopt the DARTS based NAS methods presented in Sec.~\ref{NAS_methods} to automatically learn the above discussed two sets of hyper-parameters of TDNN-F models. The over-parameterized super-network containing all possible left and right splicing context offsets and bottleneck projection dimensionality choices and the associated parameter sharing technique will be introduced in the next section.


\section{Search Space and Parameter Sharing}
This section describes the search space and its implementation when NAS methods of Sec.~\ref{NAS_methods} are used to automatically learn two sets of hyper-parameters of TDNN-F models: i) the left and right splicing context offsets; and ii) the dimensionality of the bottleneck linear projection at each hidden layer. Parameter sharing among candidate architectures used to facilitate efficient search over a large number of TDNN-F systems is also presented. Finally, NAS lattice used to extract top-ranked candidate models is introduced.

\vspace{-3mm}
\subsection{TDNN-F Context Offset Search Space}
\label{layer_context_search_space}
\begin{figure}[ht]
    \vspace{-3mm}
    \centering
    \includegraphics[width=3.5in]{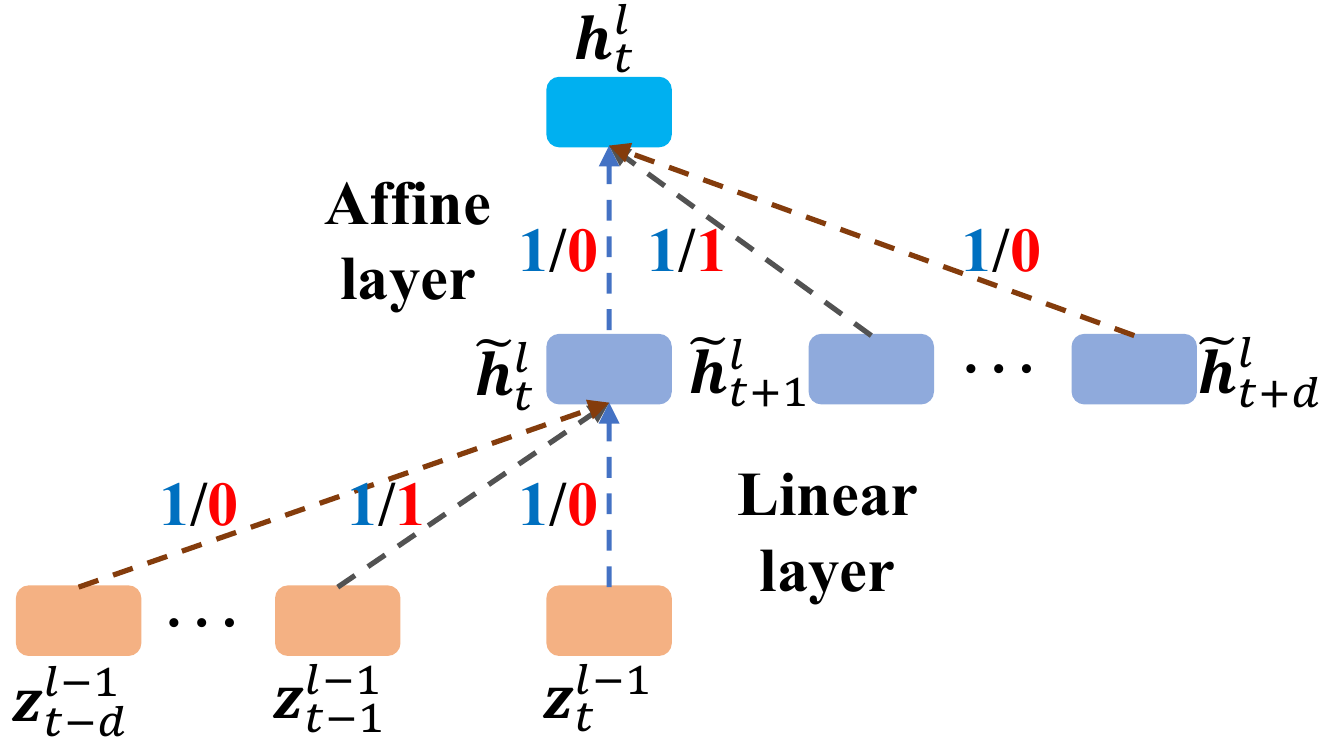}
    \vspace*{-6.5mm}
    \caption{Example part of a super-network containing all the context offsets for a TDNN-F layer. Dashed lines with different colors represent different context choices in each linear (left context) and affine (right context) transforms. The blue integers denote the super-network system using all the context offsets, while the red integers represent a candidate offset choice of $\pm1$.}
    \label{fig:layer_context_search_space}
    \vspace{-3mm}
\end{figure}
Context offset settings play an important role in modelling the long temporal information in TDNN-F models. However, manually selecting context offsets is time-consuming for different applications. Inspired by the parameter-sharing used in earlier NAS research~\cite{pham2018efficient}, we design a TDNN-F super-network (Fig.~\ref{fig:layer_context_search_space}) to contain all possible choices of context offsets to the left (\{-d,0\}, $\cdots$, \{-1,0\}, \{0,0\}) and right (\{0,0\}, \{0,1\}, $\cdots$, \{0,d\}) at each layer during search. For the super-network system, it requires the sparse context connection weights to be densely set as 1 for all context offsets. Any candidate TDNN-F model with particular context offsets out of the total $(d+1)^{2L}$ possible choices contained in the super-network is represented by setting the corresponding connection weights to be 1, while setting the others to be 0. 

\vspace{-3mm}
\subsection{TDNN-F Bottleneck Dimensionality Search Space}
\label{layer_bn_dim_search_space}
\begin{figure}[!ht]
    \vspace{-3mm}
    \centering
    \includegraphics[width=3.5in]{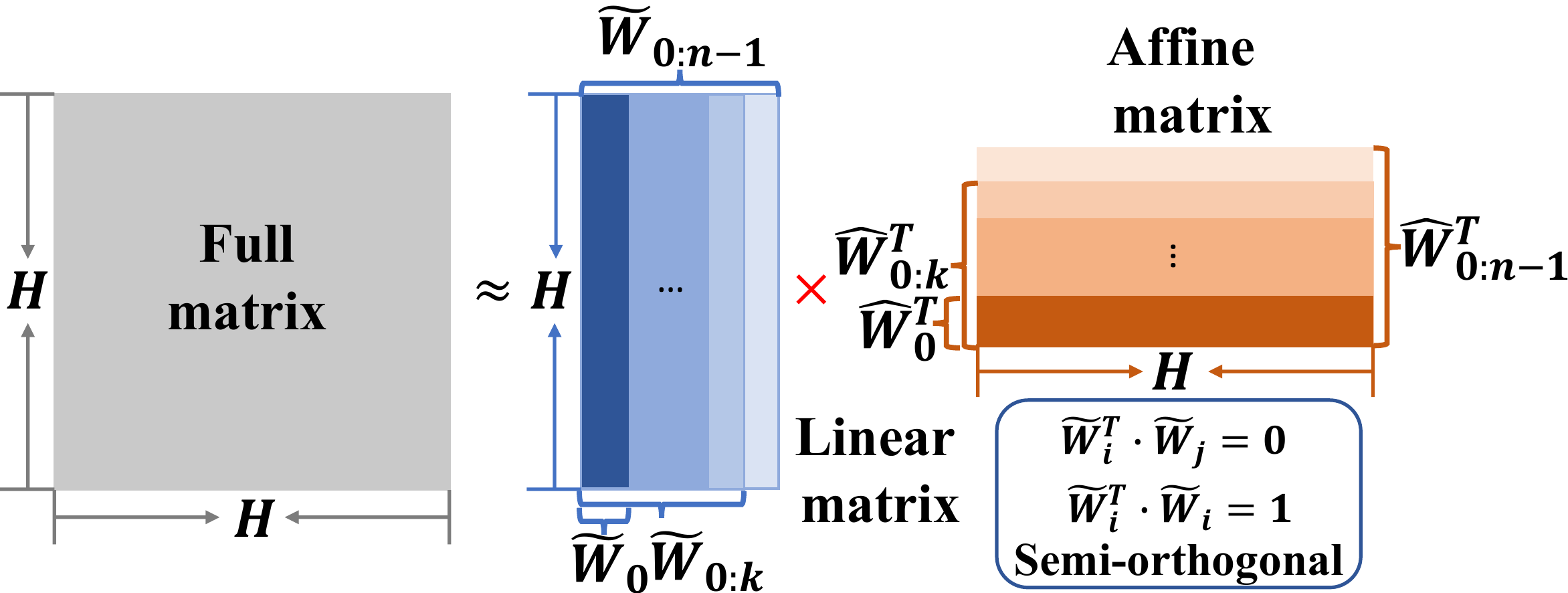}
    \vspace*{-6.5mm}
    \caption{Example part of a super-network containing different bottleneck projection dimensionality choices in the TDNN-F hidden layer. The full weight matrix is factored into one semi-orthogonal linear weight matrix $\widetilde{\mathbf{W}}_{0:n-1}$ and one affine weight matrix $\widehat{\mathbf{W}}_{0:n-1}$. Architectures with different projection dimensions are represented by the corresponding submatrices starting form the first column.}
    \label{fig:layer_bottleneck_dim_search_space}
    \vspace{-3mm}
\end{figure}
Similarly, a TDNN-F super-network containing all the candidate architectures with different projection dimensions is designed, as shown in Fig.~\ref{fig:layer_bottleneck_dim_search_space} for one hidden layer. When applying the NAS methods in Sec.2, $\phi_i^l(\cdot)$ of Eq. 1 is set as an identity matrix. In common with the standard TDNN-F model, the weight matrix $\mathbf{W}_i^{l}$ of the $i$-th architecture choice in the $l$-th layer is factored into one semi-orthogonal weight matrix $\widetilde{\mathbf{W}}^{l}_{0:n_i-1}$ and one affine weight matrix $\widehat{\mathbf{W}}^{l}_{0:n_i-1}$ as shown in Fig.~\ref{fig:layer_bottleneck_dim_search_space}. $n_i$ is the dimensionality of the $i$-th architecture. Parameter sharing among different candidate architectures' linear matrices $\widetilde{\mathbf{W}}_{0:k}$ (left to right from the first column) and affine matrices $\widehat{\mathbf{W}}_{0:k}$ (bottom to up from the first row) ($0\leq k\leq n-1$) is implemented by taking the corresponding submatrices extracted from the largest matrix $\widetilde{\mathbf{W}}_{0:n-1}$. Such sharing allows a large number of TDNN-F projection dimensionality choices at each of the 14 layers, e.g., selected from 8 values \{25, 50, 80, 100, 120, 160, 200, 240\} as considered in this paper, to be compared for selection during search. This leads to a total of $8^{14}$ candidate TDNN-F systems to be selected from.

\vspace{-3mm}
\subsection{NAS lattice}
\begin{figure}[ht]
    \vspace{-3mm}
    \centering
    \includegraphics[width=3.5in]{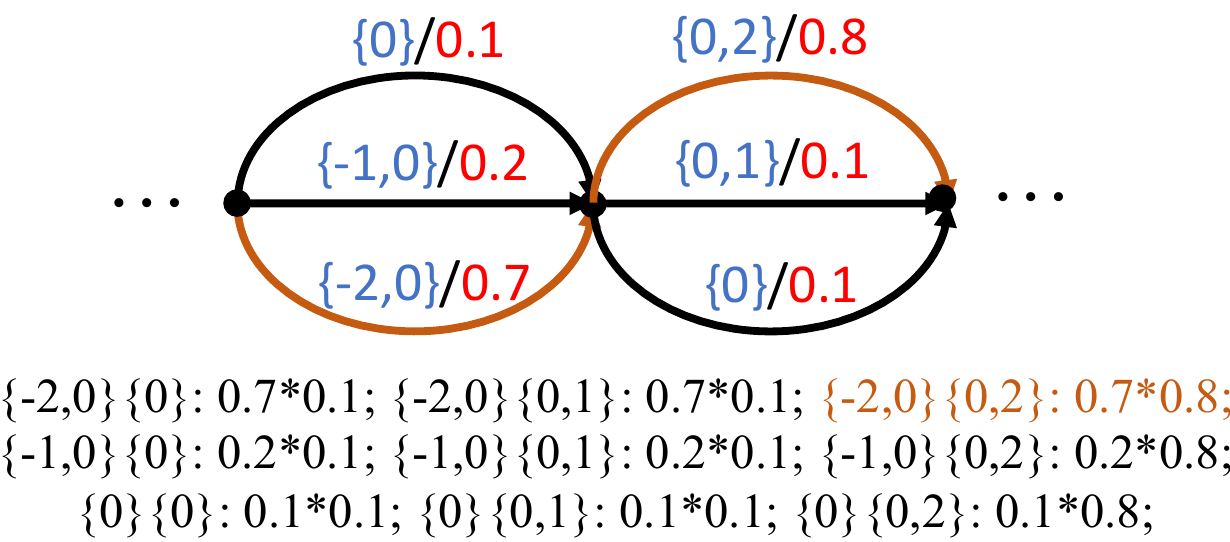}
    \vspace*{-6.5mm}
    \caption{Part of an example NAS lattice containing architecture weights. Blue integers denote different TDNN-F left or right context offset choices, while red integers are their associated weights. Among all 9 possible context choices shown in the figure, the brown colored path with +/-2 offsets is chosen with the highest probability $0.7*0.8=0.56$.}
    \label{fig:lattice_ranking}
    \vspace{-3mm}
\end{figure}
When the architecture weights are learned using various NAS methods presented in Sec.~\ref{NAS_methods}, all the candidate architectures contained in the super-network (as introduced in Sec.~\ref{layer_context_search_space} and Sec.~\ref{layer_bn_dim_search_space}) can be represented by a NAS lattice carrying their associated weights as the measure of their ranking order. Top 1-best or top N-best candidate architectures can then be extracted from the resulting NAS lattice. An example part of a NAS lattice for determining the TDNN-F left and right context offsets is shown in Fig.~\ref{fig:lattice_ranking}.

\section{Experiments}
This section is organized as follows. Firstly, the performance of learning two architecture hyper-parameters that heavily affect the performance and model complexity of state-of-the-art factored time delay neural network (TDNN-F)~\cite{waibel1989consonant, peddinti2015time, povey2018semi, povey2016purely} acoustic models trained on the benchmark 300-hour Switchboard corpus and featuring speed perturbation, i-Vector and learning hidden unit contribution (LHUC)~\cite{swietojanski2016learning} based speaker adaptation as well as RNNLM rescoring is presented in Sec.~\ref{sec: exp_nas_swbd}. Further analysis using Bayesian learning are conducted to show that the proposed NAS approaches can effectively minimize the structural redundancy in the TDNN-F systems and reduce their model parameter uncertainty. Secondly, to further evaluate the performance of the proposed NAS techniques, they were applied to automatically configure the same two sets of hyper-parameters of a state-of-the-art dysarthric speech recognition task based on the UASpeech corpus~\cite{kim2008dysarthric,liu2021recent} in Sec.~\ref{sec: exp_nas_uaspeech}. 

All of our models were trained with one thread on a single NVIDIA Tesla V100 Volta GPU card. In the searching stage, TDNN-F super-network models are trained on the training set with one thread for 3 epochs, while architecture parameters of PipeSoftmax and PipeGumbel systems are updated for additional 3 epochs using a held-out data set by fixing the normal DNN parameters. Note that we randomly select 5\% of the original training set as the held-out data set and $T$ in the Gumbel-Softmax distribution is linearly annealed from 1 to 0.03 in our experiments. Once candidate TDNN-F models are derived from the searching stage, they are trained for 3 epochs from scratch. The matched pairs sentence-segment word error (MAPSSWE) based statistical significance test~\cite{gillick1989some} was performed at a significance level $\alpha=0.05$.

\vspace{-3mm}
\subsection{NAS Experiments on 300-Hour Switchboard Task}
~\label{sec: exp_nas_swbd}
This section presents our NAS experiments of learning either the left and right splicing context offsets or bottleneck linear projection dimensionality or both of them at each layer on the 300-hour (900-hour after speed perturbation) Switchboard telephone speech recognition task using the Kaldi toolkit \cite{povey2011kaldi}. In all our experiments, we follow the Kaldi chain model setup\footnote{All of this is in published Kaldi code at https://github.com/kaldi-asr/kaldi/tree/master/egs/swbd/s5c/run.sh and https://github.com/kaldi-asr/kaldi/tree/master/egs/swbd/s5c/local/chain/tuning/run\_tdnn\_7q.sh.}, except that we used 40-dimension filterbank features for neural network training instead of the 40-dimension high-resolution Mel-frequency cepstral coefficients (MFCCs).


{\bf Task Description:} The Switchboard I telephone speech corpus consists of approximately 300 hours audio data released by LDC (LDC97S62). The baseline GMM-HMM system was trained based on 40-dimensional Mel-frequency cepstral coefficients (MFCCs) to generate alignments for the neural network training. For performance evaluation, a four-gram language model (LM) trained on the Switchboard and Fisher transcripts (LDC2004T19, LDC2005T19) was used to evaluate NIST HUB5' 00 (LDC2002S09, LDC2002T43), RT03 (LDC2007S10) and RT02 (LDC2004S11) test sets. In addition, the Kaldi recipe LSTM recurrent neural network language model (RNNLM) trained on the Switchboard and Fisher transcripts (LDC2004T19, LDC2005T19) was used to rescore the nbest lists produced by the LF-MMI trained systems with a four-gram language model (LM). The performance of LF-MMI trained TDNN baseline system incorporated with i-Vector~\cite{dehak2010front} and speed perturbation is shown in line 1 of Tab.~\ref{exp: tdnn_context_offset_results}, Tab.~\ref{exp: tdnn_bn_dim_results} and Tab.~\ref{exp: tdnn_context_offset_bn_dim_results}. Furthermore, the effects of LHUC~\cite{swietojanski2016learning} based speaker adaptation were investigated.

\begin{table*}[t]
\vspace{-2mm}
\caption{Performance (WER\%) comparison of TDNN-F models configured with context offsets produced by the baseline system, manual designed systems, Softmax DARTS (Softmax), Gumbel-Softmax DARTS (Gumbel), Pipelined Softmax DARTS (PipeSoftmax), Pipelined Gumbel-Softmax DARTS (PipeGumbel) systems described in Sec.~\ref{NAS_methods}. $\{[a,b]\}$:$\{-c, d\}$ denotes context offsets $\{-c,0\}$ to the left and $\{0,d\}$ to the right used from $a$-th layer to $b$-th layer inclusive. $^\dagger$ denotes a statistically significant difference is obtained over the baseline system (Sys (1)). (SWB1 and CHM denote the Switchboard and Callhm subsets of the Hub5' 00 test set; FSH and SWB2 denote the Fisher and Switchboard subsets of the RT03S test set; SWB3, SWB4 and SWB5 denote three Switchboard subsets in the Rt02 test set.)}
\label{exp: tdnn_context_offset_results}
\centering
\setlength{\tabcolsep}{1.2mm}{
\begin{tabular}{c|c|c|c|cc|cc|ccc|c|c|c}
\hline\hline
\multirow{2}*{Sys} & \multirow{2}*{Method} & I-Vec & \multirow{2}*{Context Offsets} & \multicolumn{2}{c|}{Hub5' 00} & \multicolumn{2}{c|}{Rt03S} & \multicolumn{3}{c|}{Rt02} & \multirow{2}*{Avg} & \multirow{2}*{\#param} & \multirow{2}*{Time}\\
\cline{5-11}
~ & ~ & +SP & ~ & SWB1 & CHM & FSH  & SWB2 & SWB3 & SWB4 & SWB5 & ~ & ~\\
\hline\hline
1 & Baseline & \cmark & \{[1,3]\}:\{-1,1\}; \{4\}:\{0\}; \{[5,14]\}:\{-3,3\} & 9.7 & 18.0 & 12.6 & 19.5 & 11.5 & 15.3 & 20.0 & 15.5 & 18.6M & 30h\\
\hline
2 & \multirow{7}*{Manual} & \multirow{7}*{\cmark} & \{1\}:\{-1,1\}; \{2\}:\{0\}; \{[3,14]\}:\{-3,3\} & 9.4$^\dagger$ & 17.9 & 12.5 & 19.4 & 11.4 & 15.3 & 19.4$^\dagger$ & 15.4 & \multirow{7}*{18.6M} & \multirow{7}*{30h}\\
3 & ~ & ~ & \{[1,2]\}:\{-1,1\}; \{3\}:\{0\}; \{[4,14]\}:\{-3,3\} & 9.4$^\dagger$ & 17.8 & 12.3$^\dagger$ & 19.6 & 11.4 & 15.0$^\dagger$ & 19.5$^\dagger$ & 15.3 & & \\
4 & ~ & ~ & \{[1,4]\}:\{-1,1\}; \{5\}:\{0\}; \{[6,14]\}:\{-3,3\} & 9.6 & 17.6$^\dagger$ & 12.4 & 19.4 & 11.3 & 15.5 & 19.4$^\dagger$ & 15.3 & & \\
5 & ~ & ~ & \{[1,7]\}:\{-1,1\}; \{8\}:\{0\}; \{[9,14]\}:\{-3,3\} & 9.4$^\dagger$ & 17.9 & 12.3$^\dagger$ & 19.6 & 11.4 & 15.4 & 19.4$^\dagger$ & 15.4 &  & \\
6 & ~ & ~ & \{[1,3]\}:\{-1,1\}; \{4\}:\{0\}; \{[5,14]\}:\{-6,6\} & 9.4$^\dagger$ & 17.5$^\dagger$ & 12.1$^\dagger$ & 19.0$^\dagger$ & 11.4 & 15.1 & 18.9$^\dagger$ & \textbf{15.1}$^\dagger$ &  & \\
7 & ~ & ~ & \{[1,3]\}:\{-1,1\}; \{4\}:\{0\}; \{[5,14]\}:\{-9,9\} & 9.2$^\dagger$ & 17.7$^\dagger$ & \textbf{11.9}$^\dagger$ & \textbf{18.7}$^\dagger$ & 11.1$^\dagger$ & 14.9$^\dagger$ & 18.7$^\dagger$ & \textbf{14.9}$^\dagger$ &  & \\
8 & ~ & ~ & \{[1,3]\}:\{-1,1\}; \{4\}:\{0\}; \{[5,14]\}:\{-12,12\} & 9.6 & 18.1 & 12.1$^\dagger$ & 18.9$^\dagger$ & 11.3 & 15.0$^\dagger$ & 18.8$^\dagger$ & 15.1$^\dagger$ &  &  \\
\hline\hline
9 & \tabincell{c}{Softmax \\Supernet} &  \multirow{11}*{\cmark} & \{[1,14]\}:\{[-6,6]\} & 9.5 & 17.6 & 12.4 & 19.7 & 11.2 & 15.0 & 19.6 & 15.4 & 53.5M & 113h\\
\cline{1-2}\cline{4-14}
10 & \tabincell{c}{Softmax \\Top1} & ~ & \tabincell{c}{\{1\}:\{0,6\}; \{[2,4]\}:\{-2,2\}; \{5\}:\{0\}; \\ \{6\}:\{0,1\}; \{7\}:\{-1,2\}; \{[8]\}:\{-2,6\}; \\ \{9\}:\{-6,3\}; \{10\}:\{-4,6\};\{[11,14]\}:\{-6,6\}} & 9.4$^\dagger$ & 17.4$^\dagger$ & 12.3$^\dagger$ & 19.4 & 11.1$^\dagger$ & 15.2 & 19.5$^\dagger$ & 15.2$^\dagger$ & 18.2M & \multirow{7}*{30h} \\
\cline{1-2}\cline{4-13}
11 & \tabincell{c}{Softmax \\Top2} & ~ & \tabincell{c}{\{1\}:\{0,6\}; \{2,4\}:\{-2,2\}; \{3\}:\{-2,3\}; \{5\}:\\ \{0\}; \{6\}:\{0,1\}; \{7\}:\{-1,2\}; \{[8]\}:\{-2,6\}; \\ \{9\}:\{-6,3\}; \{10\}:\{-4,6\};\{[11,14]\}:\{-6,6\}} & 9.4$^\dagger$ & 17.5$^\dagger$ & 12.2$^\dagger$ & 19.5 & 11.1$^\dagger$ & 15.0$^\dagger$ & 19.3$^\dagger$ & 15.2$^\dagger$ & 18.2M &  \\
\cline{1-2}\cline{4-13}
12 & \tabincell{c}{Softmax \\Top3} & ~ & \tabincell{c}{\{1\}:\{0,6\}; \{2\}:\{-2,3\}; \{[3,4]\}:\{-2,2\}; \{5\}:\\ \{0\}; \{6\}:\{0,1\}; \{7\}:\{-1,2\}; \{[8]\}:\{-2,6\}; \\ \{9\}:\{-6,3\}; \{10\}:\{-4,6\};\{[11,14]\}:\{-6,6\}} & 9.4$^\dagger$ & 17.4$^\dagger$ & 12.2$^\dagger$ & 19.3 & 11.2$^\dagger$ & 15.3 & 19.2$^\dagger$ & 15.2$^\dagger$ & 18.2M &  \\
\hline\hline
13 & \tabincell{c}{Gumbel \\ Supernet} & \multirow{10}*{\cmark} & \{[1,14]\}:\{[-6,6]\} & 10.4 & 19.3 & 13.6 & 20.9 & 12.1 & 16.3 & 20.8 & 16.6 & 53.5M & 113h\\
\cline{1-2}\cline{4-14}
14 & \tabincell{c}{Gumbel \\ Top1} & ~ & \tabincell{c}{\{1\}:\{0\}; \{2,3\}:\{-3,3\}; \{4\}:\{-4,3\}; \{5,6\}:\\ \{-4,4\}; \{7\}:\{-5,6\}; \{[8,14]\}:\{-6,6\}} & \textbf{9.3}$^\dagger$ & 17.6$^\dagger$ & 12.1$^\dagger$ & 19.3 & 11.2$^\dagger$ & 15.0$^\dagger$ & 19.0$^\dagger$ & \textbf{15.1}$^\dagger$ & 18.6M & \multirow{6}*{30h} \\
\cline{1-2}\cline{4-13}
15 & \tabincell{c}{Gumbel \\ Top2} & ~ & \tabincell{c}{\{1\}:\{0\}; \{2,4\}:\{-3,3\}; \{3\}:\{-2,3\}; \\ \{5\}:\{-4,3\}; \{6,7\}:\{-4,4\}; \\ \{8\}:\{-5,6\}; \{[9,14]\}:\{-6,6\}} & \textbf{9.3}$^\dagger$ & \textbf{16.9}$^\dagger$ & \textbf{11.9}$^\dagger$ & 19.1$^\dagger$ & 11.1$^\dagger$ & 15.0$^\dagger$ & \textbf{18.6}$^\dagger$ & \textbf{14.9}$^\dagger$ & 18.6M &  \\
\cline{1-2}\cline{4-13}
16 & \tabincell{c}{Gumbel \\ Top3} & ~ & \tabincell{c}{\{1\}:\{0,3\}; \{2\}:\{-3,3\}; \{3\}:\{-3,4\}; \\ \{4\}:\{-3,2\}; \{5,6\}:\{-4,4\}; \\ \{7\}:\{-5,6\}; \{[8,14]\}:\{-6,6\}} & 9.5 & 17.6$^\dagger$ & 12.1$^\dagger$ & 19.6 & \textbf{11.0}$^\dagger$ & 15.2 & 19.2$^\dagger$ & 15.2$^\dagger$ & 18.9M &  \\
\hline\hline
17 & \tabincell{c}{PipeSoftmax \\Supernet} & \multirow{12}*{\cmark} & \{[1,14]\}:\{[-6,6]\} & 9.7 & 18.2 & 12.5 & 19.9 & 11.3 & 14.9 & 19.6 & 15.5 & 53.6M & 80h\\
\cline{1-2}\cline{4-14}
18 & \tabincell{c}{PipeSoftmax \\Top1} & ~ & \tabincell{c}{\{1\}:\{-1,2\}; \{2\}:\{-2,2\}; \{3\}:\{-2,5\};\\ \{4\}:\{-3,6\}; \{5\}:\{-4,5\}; \{6\}:\{-5,6\};\\ \{9\}:\{-6,5\}; \{7,8,[10,14]\}:\{-6,6\}} & 9.2$^\dagger$ & 17.4$^\dagger$ & 12.2$^\dagger$ & 19.2$^\dagger$ & 11.2$^\dagger$ & 15.0$^\dagger$ & 18.7$^\dagger$ & \textbf{15.0}$^\dagger$ & 19.2M & \multirow{7}*{30h}\\
\cline{1-2}\cline{4-13}
19 & \tabincell{c}{PipeSoftmax \\Top2} & ~ & \tabincell{c}{\{1\}:\{-1,2\}; \{2\}:\{-2,2\}; \{3\}:\{-2,5\};\\ \{4\}:\{-2,6\}; \{5\}:\{-4,5\}; \{6\}:\{-5,6\};\\ \{9\}:\{-6,5\}; \{7,8,[10,14]\}:\{-6,6\}} & 9.4$^\dagger$ & 17.6$^\dagger$ & 12.0$^\dagger$ & 19.5 & 11.2$^\dagger$ & 15.0$^\dagger$ & 19.0$^\dagger$ & \textbf{15.1}$^\dagger$ & 19.2M & \\
\cline{1-2}\cline{4-13}
20 & \tabincell{c}{PipeSoftmax \\Top3} & ~ & \tabincell{c}{\{1\}:\{-1,2\}; \{2\}:\{-2,2\}; \{3\}:\{-2,5\};\\ \{4\}:\{-3,6\}; \{5\}:\{-4,5\}; \{6\}:\{-5,6\};\\ \{9\}:\{-6,5\}; \{11\}:\{-6,0\}; \\ \{7,8,10,[12,14]\}:\{-6,6\}} & 9.4$^\dagger$ & 17.5$^\dagger$ & 12.1$^\dagger$ & 19.3 & 11.1$^\dagger$ & 15.0$^\dagger$ & 19.6$^\dagger$ & 15.2$^\dagger$ & 19.2M & \\
\hline\hline
21 & \tabincell{c}{PipeGumbel \\Supernet} & \multirow{8}*{\cmark} & \{[1,14]\}:\{[-6,6]\} & 10.5 & 19.8 & 13.8 & 21.9 & 12.1 & 16.5 & 21.8 & 17.0 & 53.6M & 80h\\
\cline{1-2}\cline{4-14}
22 & \tabincell{c}{PipeGumbel \\Top1} & ~ & \tabincell{c}{\{1\}:\{-2,2\}; \{2\}:\{-2,4\}; \{3\}:\{-5,5\};\\ \{5\}:\{-6,5\}; \{4,[6,14]\}:\{-6,6\}} & \textbf{9.3}$^\dagger$ & 17.3$^\dagger$ & 12.4 & 19.1$^\dagger$ & \textbf{11.0}$^\dagger$ & 15.0$^\dagger$ & 19.0$^\dagger$ & \textbf{15.1}$^\dagger$ & 19.2M & \multirow{5}*{30h}\\
\cline{1-2}\cline{4-13}
23 & \tabincell{c}{PipeGumbel \\Top2} & ~ & \tabincell{c}{\{1\}:\{-2,2\}; \{2\}:\{-2,5\}; \{3\}:\{-5,5\};\\ \{5\}:\{-6,5\}; \{4,[6,14]\}:\{-6,6\}} & 9.4$^\dagger$ & 17.6$^\dagger$ & 12.2$^\dagger$ & 19.3 & 11.3 & \textbf{14.8}$^\dagger$ & 18.9$^\dagger$ & \textbf{15.1}$^\dagger$ & 19.2M & \\
\cline{1-2}\cline{4-13}
24 & \tabincell{c}{PipeGumbel \\Top3} & ~ & \tabincell{c}{\{1\}:\{-2,2\}; \{2\}:\{-2,4\}; \{3\}:\{-5,5\}; \\ \{[4,14]\}:\{-6,6\}} & 9.4$^\dagger$ & 17.9 & 12.3$^\dagger$ & 19.3 & 11.4 & 14.9$^\dagger$ & 19.3$^\dagger$ & 15.2$^\dagger$ & 19.2M & \\
\hline\hline
\end{tabular}}
\vspace{-2mm}
\end{table*}

\subsubsection{\textbf{TDNN-F Context Offset Search}}
In this section, we describe the experimental results of searching context offsets at each factored TDNN layer by using various NAS methods of Sec.~\ref{NAS_methods}. In Tab.~\ref{exp: tdnn_context_offset_results}, system (1) is the baseline Kaldi recipe$^1$ trained factored TDNN system. As a direct and exhaustive search over all possible context offset settings is infeasible, a heuristic based two stage manual search is adopted. First, the position of the single hidden layer where the left and right context offsets are both restricted to 0 (default position 4th hidden layer in Sys (1)), was relocated to the 2nd, 3rd, 5th and 8th hidden layers respectively. These changes produced manual systems (2)-(5) shown in Tab.~\ref{exp: tdnn_context_offset_results}, and were designed to intuitively vary the ratio between the hidden layers in the lower section of the system modelling shorter range of temporal contexts, and those in the higher section capturing longer span information. No significant performance difference can be obtained over the baseline recipe setting (Sys (1)). Second, this is then followed by fixing the L/R contexts as 0 in the 4th layer while further enlarging those used by the higher positioned hidden layers (from layer 5 to 14) incrementally to $\pm6$, $\pm9$, $\pm12$. Among the resulting manually configured systems (6)-(8), the setting of $\pm9$ produced absolute WER reduction up to 0.6\% (Sys (7) in Tab.~\ref{exp: tdnn_context_offset_results}) across three test sets over the baseline recipe system (1). Based on these results, the following NAS experiments were conducted to perform the search over $7^{28}$ TDNN-F configurations with maximum context offsets set as $\pm6$\footnote{Systems performing the search over $10^{28}$ TDNN-F choices with the maximum context offsets of $\pm9$ can not produce better results in practice than those with the maximum context offsets of $\pm6$. Hence, all the following NAS experiments perform the search over $7^{28}$ TDNN-F choices with the maximum context offsets of $\pm6$.} at each layer. The remaining part of Tab.~\ref{exp: tdnn_context_offset_results} shows the performance and corresponding hidden layer context offsets learned using the four DARTS based NAS methods of Sec.~\ref{NAS_methods}, Softmax, Gumble-softmax (Gumble), Pipelined Softmax (PipeSoftmax) and Pipelined Gumbel-softmax (PipeGumbel). For each of these four techniques, the associated super-network system that averaging over all possible system configurations, and the top 1, 2 and 3 architecture after system retraining are presented in turn, for example, shown in system (9)-(12) for the Softmax DARTS method. Similar experiments were conducted and shown for the other three approaches in the remaining 3 parts of Tab.~\ref{exp: tdnn_context_offset_results} from system (13) to (24). Three main trends can be observed from the results in Tab.~\ref{exp: tdnn_context_offset_results}.

\begin{table*}
\vspace{-2mm}
\caption{Performance (WER\%, number of parameters) comparison of TDNN-F models configured with projection dimensions produced by the baseline system, manual designed systems, Softmax DARTS (Softmax), Gumbel-Softmax DARTS (Gumbel), Pipelined Softmax DARTS (PipeSoftmax), Pipelined Gumbel-Softmax DARTS (PipeGumbel) systems in Sec.~\ref{NAS_methods}. $\eta$ is the penalty factor in Eqn.~(\ref{eq:penalized_darts}). The dimensionality index denotes the index of 8 dimensionality choices: \{25,50,80,100,120,160,200,240\}. $^\dagger$ denotes a statistically significant difference is obtained over the baseline system (Sys (1)).}
\label{exp: tdnn_bn_dim_results}
\centering
\setlength{\abovecaptionskip}{-0cm}
\setlength{\tabcolsep}{1.6mm}{
\begin{tabular}{c|c|c|c|c|cc|cc|ccc|c|c|c}
\hline\hline
\multirow{2}*{Sys} & \multirow{2}*{Method} & I-Vec & \multirow{2}*{$\eta$} & \multirow{2}*{Bottleneck Dim Index} & \multicolumn{2}{c|}{Hub5' 00} & \multicolumn{2}{c|}{Rt03S} & \multicolumn{3}{c|}{Rt02} & \multirow{2}*{Avg} & \multirow{2}*{\#param} & \multirow{2}*{Time}\\
\cline{6-12}
~ & ~ & +SP & ~ & ~ & SWB1 & CHM & FSH  & SWB2 & SWB3 & SWB4 & SWB5 & ~ & ~\\
\hline\hline
1 & Baseline & \cmark & - & 5 5 5 5 5 5 5 5 5 5 5 5 5 5 & 9.7 & 18.0 & 12.6 & 19.5 & 11.5 & 15.3 & 20.0 & 15.5 & 18.6M & 30h\\
\hline
2 & \multirow{7}*{Manual} & \multirow{7}*{\cmark} & \multirow{7}*{-} & 0 0 0 0 0 0 0 0 0 0 0 0 0 0 & 10.0 & 18.8 & 13.1 & 20.7 & 11.9 & 15.9 & 20.9 & 16.3 & 7.5M & 22h\\
3 & ~ & ~ & ~ & 1 1 1 1 1 1 1 1 1 1 1 1 1 1 & 9.9 & 18.1 & 12.3$^\dagger$ & 20.0 & 11.6 & 15.6 & 21.0 & 15.8 & 9.6M & 24h\\
4 & ~ & ~ & ~ & 2 2 2 2 2 2 2 2 2 2 2 2 2 2 & 9.5 & 17.8 & 12.0$^\dagger$ & 19.6 & 11.5 & 15.0$^\dagger$ & 20.4 & 15.4 & 12.1M & 26h\\
5 & ~ & ~ & ~ & 3 3 3 3 3 3 3 3 3 3 3 3 3 3 & 9.5 & 17.5$^\dagger$ & 12.0$^\dagger$ & 19.4 & 11.3 & 14.8$^\dagger$ & 20.4 & 15.3 & 13.7M & 27h\\
6 & ~ & ~ & ~ & 4 4 4 4 4 4 4 4 4 4 4 4 4 4 & 9.6 & 17.8 & 12.1$^\dagger$ & 19.3 & 11.7 & 15.2 & 20.3 & 15.4 & 15.4M & 28h\\
7 & ~ & ~ & ~ & 6 6 6 6 6 6 6 6 6 6 6 6 6 6 & 9.8 & 18.2 & 12.7 & 19.9 & 12.1 & 16.0 & 20.5 & 15.9 & 22.0M & 32h\\
8 & ~ & ~ & ~ & 7 7 7 7 7 7 7 7 7 7 7 7 7 7 & 10.0 & 18.6 & 12.6 & 19.8 & 12.1 & 15.8 & 20.5 & 15.9 & 25.3M & 34h\\
\hline\hline
9 & \tabincell{c}{Softmax \\Supernet} &  \multirow{8}*{\cmark} & \multirow{8}*{0} & \{[1, 14]\}:\{0,1,2,3,4,5,6,7\} & 9.6 & 18.0 & 12.2 & 19.3 & 11.5 & 15.2 & 20.0 & 15.4 & 25.3M & 45h\\
\cline{1-2}\cline{5-15}
10 & \tabincell{c}{Softmax \\Top1} & ~ &  & 0 0 0 0 7 7 0 7 7 7 7 7 7 7& 9.6 & 17.7$^\dagger$ & 12.2$^\dagger$ & 19.4 & 11.6 & 15.3 & 20.2 & 15.4 & 19.4M & 36h\\
\cline{1-2}
11 & \tabincell{c}{Softmax \\Top2} & ~ &  & 0 0 0 0 7 0 0 7 7 7 7 7 7 7 & 9.6 & 17.5$^\dagger$ & 12.2$^\dagger$ & 19.7 & 11.1$^\dagger$ & 15.1 & 19.9 & 15.4 & 18.1M & 24h\\
\cline{1-2}
12 & \tabincell{c}{Softmax \\Top3} & ~ &  & 0 0 7 0 7 7 0 7 7 7 7 7 7 7 & 9.4$^\dagger$ & 17.6$^\dagger$ & 12.0$^\dagger$ & 19.5 & 11.3 & 15.2 & 19.9 & 15.3 & 20.7M & 24h\\
\hline\hline
13 & \tabincell{c}{Gumbel \\ Supernet} & \multirow{8}*{\cmark} & \multirow{8}*{0} & \{[1, 14]\}:\{0,1,2,3,4,5,6,7\} & 10.0 & 18.5 & 13.1 & 20.3 & 11.9 & 15.8 & 20.4 & 16.1 & 25.3M & 45h\\
\cline{1-2}\cline{5-15}
14 & \tabincell{c}{Gumbel \\Top1} & ~ & ~ & 2 3 2 0 4 5 5 6 5 5 6 5 5 5 & 9.2$^\dagger$ & \textbf{17.3}$^\dagger$ & 12.0$^\dagger$ & \textbf{19.1}$^\dagger$ & \textbf{10.7}$^\dagger$ & 14.8$^\dagger$ & 19.3$^\dagger$ & \textbf{15.0}$^\dagger$ & 17.2M & 30h\\
\cline{1-2}
15 & \tabincell{c}{Gumbel \\Top2} & ~ &  & 2 3 2 0 4 5 5 6 6 5 6 5 5 5 & 9.3$^\dagger$ & 17.4$^\dagger$ & \textbf{11.8}$^\dagger$ & 19.3 & 10.8$^\dagger$ & 14.6$^\dagger$ & \textbf{19.1}$^\dagger$ & \textbf{15.0}$^\dagger$ & 17.4M & 30h\\
\cline{1-2}
16 & \tabincell{c}{Gumbel \\Top3} & ~ &  & 2 3 2 0 4 5 5 6 5 6 6 5 5 5 & \textbf{9.1}$^\dagger$ & 17.4$^\dagger$ & 11.9$^\dagger$ & \textbf{19.1}$^\dagger$ & 10.8$^\dagger$ & 14.7$^\dagger$ & 19.3$^\dagger$ & \textbf{15.0}$^\dagger$ & 17.4M & 30h\\
\hline\hline
17 & \tabincell{c}{PipeSoftmax \\Supernet} & \multirow{8}*{\cmark} & \multirow{8}*{0} & \{[1, 14]\}:\{0,1,2,3,4,5,6,7\} & 9.7 & 18.2 & 12.6 & 20.4 & 11.4 & 14.6 & 20.0 & 15.7 & 25.3M & 45h\\
\cline{1-2}
18 & \tabincell{c}{PipeSoftmax \\Top1} & ~ & ~ & 6 6 6 6 6 5 4 0 0 0 0 0 0 7 & 9.5 & 18.3 & 12.5 & 19.7 & 11.5 & 15.4 & 19.8 & 15.6 & 15.1M & 27h\\
\cline{1-2}
19 & \tabincell{c}{PipeSoftmax \\Top2} & ~ & & 6 6 6 6 6 5 6 0 0 0 0 0 0 7 & 9.7 & 18.1 & 12.6 & 20.0 & 11.7 & 15.4 & 20.1 & 15.7 & 15.6M & 26h\\
\cline{1-2}
20 & \tabincell{c}{PipeSoftmax \\Top3} & ~ & & 6 6 6 6 6 5 5 0 0 0 0 0 0 7 & 9.7 & 18.5 & 12.8 & 20.1 & 11.7 & 15.5 & 20.6 & 15.9 & 15.3M& 27h\\
\hline
21 & \tabincell{c}{PipeSoftmax \\Top1} & \cmark & 0.3 & 2 2 2 0 2 2 2 2 2 0 0 0 0 5& 9.5 & 17.8 & 12.2$^\dagger$ & 19.6 & 11.4 & 14.9$^\dagger$ & 20.3 & 15.4 & 11.0M & 24h\\
22 & \tabincell{c}{PipeSoftmax \\Top1} & \cmark & 0.5 & 1 1 2 0 1 1 1 1 1 1 0 0 0 4 & 9.8 & 17.8 & 12.3$^\dagger$ & 20.1 & 11.3 & 15.0$^\dagger$ & 20.3 & 15.6 & 9.6M & 23h \\
\hline\hline
23 & \tabincell{c}{PipeGumbel \\Supernet} & \multirow{8}*{\cmark} & \multirow{8}*{0} & \{[1, 14]\}:\{0,1,2,3,4,5,6,7\} & 9.9 & 18.2 & 12.8 & 20.7 & 11.3 & 14.8 & 20.4 & 15.9 & 25.3M & 45h \\
\cline{1-2}\cline{5-15}
24 & \tabincell{c}{PipeGumbel \\Top1} & ~ & ~ & 6 6 6 6 5 5 1 1 2 0 0 7 7 7 & 9.5 & 17.7$^\dagger$ & 12.3$^\dagger$ & 19.3 & 11.3 & 15.0$^\dagger$ & 19.7$^\dagger$ & 15.3 & 17.5M & 30h \\
\cline{1-2}
25 & \tabincell{c}{PipeGumbel \\Top2} & ~ & & 6 6 7 6 5 5 1 1 2 0 0 7 7 7 & 9.4$^\dagger$ & 17.5$^\dagger$ & 12.3$^\dagger$ & 19.2$^\dagger$ & 11.5 & 15.2 & 19.4$^\dagger$ & 15.2$^\dagger$ & 17.8M & 30h\\
\cline{1-2}
26 & \tabincell{c}{PipeGumbel \\Top3} & ~ &  & 6 6 6 6 5 5 1 1 1 0 0 7 7 7 & 9.5 & 17.9 & 12.3$^\dagger$ & 19.2$^\dagger$ & 11.3 & 15.1 & 19.2$^\dagger$ & 15.2$^\dagger$ & 17.3M & 30h\\
\hline
27 & \tabincell{c}{PipeGumbel \\Top1} & \cmark & 0.1 & 2 3 2 0 2 2 2 1 2 1 0 3 5 5 & 9.2$^\dagger$ & 17.4$^\dagger$ & 11.9$^\dagger$ & 19.2$^\dagger$ & 10.9$^\dagger$ & \textbf{14.4}$^\dagger$ & 19.5$^\dagger$ & \textbf{15.0}$^\dagger$ & 12.4M & 26h \\
28 & \tabincell{c}{PipeGumbel \\Top1} & \cmark & 1.0 & 1 1 1 0 1 1 1 1 1 1 1 1 2 3 & 9.4$^\dagger$ & 17.8 & 12.3$^\dagger$ & 19.7 & 11.0$^\dagger$ & 14.9$^\dagger$ & 19.9 & 15.4 & 10.0M & 24h \\
\hline\hline
\end{tabular}}
\vspace{-2mm}
\end{table*}

\begin{enumerate}
    \item NAS configured TDNN-F systems (Sys (10), (14), (18), (22)) consistently outperform the baseline Kaldi recipe~\cite{povey2011kaldi} TDNN-F system (Sys (1)). For example, PipeGumbel DARTS system (Sys (22)) significantly outperforms the baseline system (Sys (1)) by 0.7\% and 1.0\% absolute WER reductions on the \textbf{CHM} subset of \textbf{Hub5' 00} test set and \textbf{SWB5} subset of \textbf{Rt02} test set. 
    \item Compared with the manually designed systems (Sys (2)-(8)) constructed using a total of 210 GPU hours (30*7 hours for 7 systems), the NAS configured TDNN-F systems (Sys (18) \& (22)) produce comparable performance to the best manually crafted system (Sys (7)) and a much smaller number of GPU hours of 110 (reduced by 48\% relative) during both architecture search and subsequent model retraining.
    \item No significant WER difference averaged over all test sets is obtained among the top 1-best Softmax (Sys (10)), Gumbel-Softmax DARTS (Sys (14)), PipeSoftmax (Sys (18)) and PipeGumbel DARTS (Sys (22)) systems, while the super-network training for pipelined DARTS systems (Sys (17) \& (21)) were in practice found to converge faster than those of the non-pipelined counterparts (Sys (9) \& (13)).
\end{enumerate}

\subsubsection{\textbf{TDNN-F Bottleneck Projection Dimensionality Search}}
To further evaluate the performance of DARTS based NAS methods of Sec.~\ref{NAS_methods}, a set of experiments comparable to those in Tab.~\ref{exp: tdnn_context_offset_results} are conducted to learn the suitable bottleneck projection dimensionality choices at each factored TDNN layer. In Tab.~\ref{exp: tdnn_bn_dim_results}, system (1) is shown as the baseline Kaldi recipe$^1$ trained TDNN-F system. Systems (2)-(8) are the manually designed factored TDNN-F systems by uniformly setting all the layer specific bottleneck projection dimensions to be one of the following values: 25, 50, 80, 100, 120, 200 or 240. 

Tab.~\ref{exp: tdnn_bn_dim_results} shows the performance of NAS based auto-configuration of bottleneck projection dimensions at each factored TDNN layer from the following choices: 25, 50, 80, 100, 120, 160, 200 or 240, using the four methods of Sec.~\ref{NAS_methods}. This set leads to a total of $8^{14}$ factored TDNN systems to be selected from. Again in common with the context offset NAS experiments of Tab.~\ref{exp: tdnn_context_offset_results}, for each of these four techniques, Softmax, Gumble-softmax (Gumble), Pipelined Softmax (PipeSoftmax) and Pipelined Gumbel-softmax (PipeGumbel), the performance of the associated super-network system (Sys (9), (13), (17) \& (23) in Tab.~\ref{exp: tdnn_bn_dim_results}), and the respective top 1, 2 and 3 architecture based systems retraining (Sys (10)-(12), (14)-(16), (18)-(22) and (24)-(28)) are presented in Tab.~\ref{exp: tdnn_bn_dim_results}. Several trends can be observed from the results in Tab.~\ref{exp: tdnn_bn_dim_results}.

\begin{table*}[t]
\vspace{-2mm}
\caption{Performance (WER\%, number of parameters) of TDNN-F baseline systems, TDNN-F models configured with both context offsets and projection dimensionality produced by Pipelined Softmax DARTS (PipeSoftmax) or Pipelined Gumbel-Softmax DARTS (PipeGumbel) or Random Search before and after applying LHUC speaker adaptation and RNNLM rescoring. Random Search selects the best performed model from 6 randomly sampled models. $\eta$ is the penalty factor in Eqn.~(\ref{eq:penalized_darts}). "Two Stage" denotes the context offsets and bottleneck projection dimensions are searched in a sequential order. "Joint" denotes context offsets and bottleneck projection dimensions are searched in the same super-network. $^\dagger$ denotes a statistically significant difference is obtained over the baseline Kaldi recipe systems (Sys (1) \& (8)).}
\label{exp: tdnn_context_offset_bn_dim_results}
\centering
\setlength{\tabcolsep}{1.6mm}{
\begin{tabular}{c|c|c|c|c|c|c|cc|cc|ccc|c|c|c}
\hline\hline
\multirow{2}*{Sys} & \multirow{2}*{Method} & I-Vec &
\multirow{2}*{LHUC} & \multirow{2}*{LM} & Context & \multirow{2}*{$\eta$} & \multicolumn{2}{c|}{Hub5' 00} &  \multicolumn{2}{c|}{Rt03S} & \multicolumn{3}{c|}{Rt02} & \multirow{2}*{Avg} & \multirow{2}*{\#param} & \multirow{2}*{Time}\\
\cline{8-14}
~ & ~ & +SP & ~ & ~ & \& Dim & ~ & SWB1 & CHM & FSH  & SWB2 & SWB3 & SWB4 & SWB5 & ~ & ~\\
\hline\hline
1 & Baseline & \multirow{12}*{\cmark} & \multirow{12}*{\xmark} & \multirow{12}*{4-gram} & \multirow{4}*{-} & \multirow{4}*{-} & 9.7 & 18.0 & 12.6 & 19.5 & 11.5 & 15.3 & 20.0 & 15.5 & 18.6M & 30h\\
\cline{1-2}\cline{8-17}
2 & Manual &  &  & &   &  & \textbf{9.1}$^\dagger$ & \textbf{17.3}$^\dagger$ & 11.9$^\dagger$ & \textbf{18.7}$^\dagger$ & \textbf{10.6}$^\dagger$ & \textbf{14.6}$^\dagger$ & 19.0$^\dagger$ & \textbf{14.8}$^\dagger$ & 13.7M & 463h\\
\cline{1-2}\cline{8-17}
3 & \tabincell{c}{Random \\Search}&   &  &  &   & 
& 9.5 & 18.2 & 12.3$^\dagger$ & 19.7 & 11.0$^\dagger$ & 15.0$^\dagger$ & 20.1 & 15.3 & 13.8M & 162h\\
\cline{1-2}\cline{6-17}
4 & \tabincell{c}{PipeSoftmax \\(Top1)} &   &  & & \multirow{3}*{\tabincell{c}{Two \\Stage}} & 0.1 & 9.5 & 17.4$^\dagger$ & 12.2$^\dagger$ & 19.5 & 11.2$^\dagger$ & 15.0$^\dagger$ & 19.5$^\dagger$ & 15.2$^\dagger$ & 13.7M & \multirow{7}*{160h}\\
\cline{1-2}\cline{7-16}
5 & \tabincell{c}{PipeGumbel \\(Top1)} &   &  &  &  & 0.3 & 9.2$^\dagger$ & \textbf{17.3}$^\dagger$ & \textbf{11.8}$^\dagger$ & 18.8$^\dagger$ & 10.7$^\dagger$ & 14.7$^\dagger$ & \textbf{18.8}$^\dagger$ & \textbf{14.8}$^\dagger$ & 12.9M & \\
\cline{1-2}\cline{6-16}
6 & \tabincell{c}{PipeSoftmax \\(Top1)} &   &  &  & \multirow{3}*{\tabincell{c}{Joint}} & 0.1 & 9.4$^\dagger$ & 17.5$^\dagger$ & 12.1$^\dagger$ & 19.4 & 11.2$^\dagger$ & 14.8$^\dagger$ & 19.5$^\dagger$ &  15.2$^\dagger$ & 11.7M & \\
\cline{1-2}\cline{7-16}
7 & \tabincell{c}{PipeGumbel \\(Top1)} &   &  &  &  & 0.3 & 9.3$^\dagger$ & 17.6$^\dagger$ & 12.1$^\dagger$ & 19.2$^\dagger$ & 11.2$^\dagger$ & 15.2 & 19.2$^\dagger$ & 15.1$^\dagger$ & 11.7M & \\
\hline\hline
8 & Baseline & \multirow{12}*{\cmark} & \multirow{12}*{\cmark} & \multirow{12}*{+RNNLM} & \multirow{4}*{-} & \multirow{4}*{-} & 7.9 & 15.2 & 10.1 & 16.3 & 9.5 & 12.4 & 16.1 & 12.8 & 18.6M & 32h\\
\cline{1-2}\cline{8-17}
9 & Manual &  &  &  &  &  & \textbf{7.4}$^\dagger$ & 14.4$^\dagger$ & 9.5$^\dagger$ & \textbf{15.3}$^\dagger$ & \textbf{8.7}$^\dagger$ & \textbf{11.8}$^\dagger$ & 15.2$^\dagger$ & \textbf{12.0}$^\dagger$ & 13.7M & 465h\\
\cline{1-2}\cline{8-17}
10 & \tabincell{c}{Random \\Search}&   &  &  &  &
& 7.6$^\dagger$ & 15.1 & 10.0 & 16.3 & 9.2$^\dagger$ & 12.3 & 16.3 & 12.7 & 13.8M & 164h\\
\cline{1-2}\cline{6-17}
11 & \tabincell{c}{PipeSoftmax \\(Top1)} &   &  &  & \multirow{3}*{\tabincell{c}{Two \\Stage}}  & 0.1 & 7.6$^\dagger$ & 14.6$^\dagger$ & 9.8$^\dagger$ & 15.9$^\dagger$ & 9.1$^\dagger$ & 11.9$^\dagger$ & 15.9 & 12.3$^\dagger$ & 13.7M & \multirow{7}*{162h}\\
\cline{1-2}\cline{7-16}
12 & \tabincell{c}{PipeGumbel \\(Top1)} &   &  &  &  & 0.3  & \textbf{7.4}$^\dagger$ & \textbf{14.2}$^\dagger$ & \textbf{9.4}$^\dagger$ & 15.4$^\dagger$ & 8.8$^\dagger$ & 11.9$^\dagger$ & 15.1$^\dagger$ & \textbf{12.0}$^\dagger$ & 12.9M & \\
\cline{1-2}\cline{6-16}
13 & \tabincell{c}{PipeSoftmax \\(Top1)} &   &  &  & \multirow{3}*{\tabincell{c}{Joint}} & 0.1 & 7.6$^\dagger$ & 14.3$^\dagger$ & 9.5$^\dagger$ & 15.8$^\dagger$ & \textbf{8.7}$^\dagger$ & \textbf{11.8}$^\dagger$ & \textbf{15.0}$^\dagger$ & 12.1$^\dagger$ & 11.7M & \\
\cline{1-2}\cline{7-16}
14 & \tabincell{c}{PipeGumbel \\(Top1)} &   &  &  & & 0.3 & 7.7 & 14.7$^\dagger$ & 9.6$^\dagger$ & 15.7$^\dagger$ & 9.1$^\dagger$ & 12.2 & 15.3$^\dagger$ & 12.3$^\dagger$ & 11.7M & \\
\hline\hline
\end{tabular}}
\end{table*}

\begin{enumerate}
    \item When compared with the baseline Kaldi recipe~\cite{povey2011kaldi} system (Sys (1)), the Gumbel and PipeGumbel systems (Sys (14), (24)) with a similar model size of approximately 17 million parameters achieve comparable or better performance. For example, the Gumbel Top 1 system (Sys (14)) produces 0.5\% absolute significant WER reduction on average across all test sets.
    \item When compared with the best-performing manually designed system (Sys (5)), Gumbel Top 1 system (Sys (14)) also produces 0.3\% statistically significant absolute WER reduction on average across all three test sets.
    \item If we further add the resource penalty to the objective loss function as presented in the penalized DARTS method of Sec.~\ref{sec: penalized_darts}, the PipeGumbel system (Sys (27), with penalty coefficient $\eta=0.1$) can produce statistically significant absolute WER reductions ranging from 0.3\% (\textbf{SWB2} subset of \textbf{Rt03S} test set) to 0.9\% (\textbf{SWB4} subset of \textbf{Rt02} test set) and a relative model size reduction of 33\% over the baseline Kaldi recipe trained TDNN-F system (Sys (1)), by selecting fewer bottleneck projection dimensions at higher layers than the PipeGumbel system (Sys (24)) using no model size penalty.
    \item The PipeGumbel system (Sys (28) with penalty coefficient $\eta=1.0$) achieves the largest relative model size reduction of 46\% with a marginal WER reduction of 0.1\% when compared with the baseline recipe TDNN-F system (Sys (1)). This serves an alternative penalized PipeGumbel system setting to system (27), which favours the most aggressively compressed model architecture incurring no accuracy performance degradation.
\end{enumerate}

Note that the resource penalty was not added to the architecture search stage of Softmax and Gumbel-Softmax DARTS systems due to the requirement of retraining super-network parameters for different resource penalty scaling factors. For efficiency, in the following NAS experiments where both the hidden layer L/R context offsets and bottleneck projection dimensions are automatically determined, only Pipelined Softmax and Gumble-Softmax DARTS systems are used.


\begin{table*}[h!]
\caption{Performance (WER\%, number of parameters) comparison of TDNN-F systems combined with Bayesian estimation or Pipelined Gumbel DARTS (PipeGumbel) or both of them. "Parameter Variance" denotes the average parameter latent distribution's variance in the Bayesian estimated factored TDNN systems. $^\dagger$ denotes a statistically significant difference is obtained by NAS configured TDNN systems and Bayesian estimated TDNN systems (line 2-4) over the TDNN baseline system (line 1).}
\label{exp: tdnn_nas_bayesian}
\centering
\setlength{\tabcolsep}{1.6mm}{
\begin{tabular}{c|c|c|c|c|c|cc|cc|ccc|c|c|c}
\hline\hline
 &
\multirow{2}*{System} &
I-Vec &
\multirow{2}*{LHUC} & \multirow{2}*{LM} & Context Offset \& & \multicolumn{2}{c}{Hub5' 00} & \multicolumn{2}{c}{Rt03S} & \multicolumn{3}{c|}{Rt02} & \multirow{2}*{Avg} & \multirow{2}*{\#Param} & Parameter\\
\cline{7-13}
~ & ~ & +SP & ~ & &  Bottleneck Dim & SWB1 & CHM & FSH  & SWB2 & SWB3 & SWB4 & SWB5 & & & Variance\\
\hline\hline
1 & \multirow{2}*{TDNN} & \multirow{4}*{\cmark} & \multirow{4}*{\xmark} & \multirow{4}*{4-gram} & Manual & 9.7 & 18.0 & 12.6 & 19.5 & 11.5 & 15.3 & 20.0 & 15.5 & 18.6M & -\\
2 &  &  &  &  & NAS & 9.2$^\dagger$ & 17.3$^\dagger$ & 11.8$^\dagger$ & 18.8$^\dagger$ & 10.7$^\dagger$ & 14.7$^\dagger$ & 18.8$^\dagger$ & 14.8$^\dagger$ & 12.9M & -\\
\cline{2-2}\cline{6-16}
3 & Bayes &  &  &  & Manual & 9.4$^\dagger$ & 17.3$^\dagger$ & 12.1$^\dagger$ & 19.2$^\dagger$ & 11.4 & 14.7$^\dagger$ & 19.1$^\dagger$ & 15.1$^\dagger$ & 18.6M & 0.033\\
4 & TDNN &  &  &  & NAS & 9.3$^\dagger$ & 17.5$^\dagger$ & 11.8$^\dagger$ & 18.8$^\dagger$ & 11.1$^\dagger$ & 14.7$^\dagger$ & 18.6$^\dagger$ & 14.8$^\dagger$ & 12.9M & 0.025\\
\hline\hline
\end{tabular}}
\vspace{-2mm}
\end{table*}

\begin{table*}[h]
\caption{Performance (WER\%, number of parameters) of TDNN-F models configured with both context offsets and projection dimensions produced by NAS using the original 300-hour data and the 900-hour augmented data.}
\vspace{-2mm}
\centering
\label{exp: swbd_nas_transferability_result}
\begin{tabular}{c|c|c|c|c|c|cc|cc|ccc|c|c}
\hline\hline
\multirow{2}*{Sys} & \multirow{2}*{Method} & \multirow{2}*{Search} & \multirow{2}*{Retrain} &
\multirow{2}*{LHUC} & \multirow{2}*{LM} & \multicolumn{2}{c|}{Hub5' 00} & \multicolumn{2}{c|}{Rt03S} & \multicolumn{3}{c|}{Rt02} & \multirow{2}*{Avg} & \multirow{2}*{\#param} \\
\cline{7-13}
~ & ~ &  & ~ & &  & SWB1 & CHM & FSH  & SWB2 & SWB3 & SWB4 & SWB5 & ~ & ~\\
\hline\hline
1 & \multirow{3}*{NAS} & 300h & 300h & \multirow{3}*{\xmark} & \multirow{3}*{4g} & 9.6 & 19.8 & 13.2 & 20.9 & 11.6 & 15.6 & 21.8 & 16.4 & 13M\\
\cline{1-1}\cline{7-15}
2 &  & 300h & 900h &  &  & 9.5 & 17.3 & 12.2 & 19.7 & 11.4 & 15.1 & 19.6 & 15.3 & 13M\\
\cline{1-1}\cline{7-15}
3 &  & 900h & 900h & & & 9.2 & 17.3 & 11.8 & 18.8 & 10.7 & 14.7 & 18.8 & 14.8 & 13M\\
\hline\hline
\end{tabular}
\vspace{-2mm}
\end{table*}

\begin{table*}[h]
\vspace{-2mm}
\caption{Performance contrasts of TDNN-F/CNN-TDNN-F models configured with both context offsets and projection dimensionality produced by baseline system or Pipelined Gumbel (PipeGumbel) DARTS (after applying Bayesian LHUC and large RNNLMs) against other state-of-the-art systems conducted on the 300-hour Switchboard task. The overall WERs in "()" are not reported by the original papers and are recalculated using the subset WERs.}
\label{exp: swbd_nas_newest_result}
\centering
\begin{tabular}{c|l|c|ccc|ccc}
\hline\hline
 &
\multirow{2}*{System} &
\multirow{2}*{\#Param} & \multicolumn{3}{c}{Hub5' 00} & \multicolumn{3}{c}{Rt03S}\\
\cline{4-9}
~ & ~ & ~ & SWB1 & CHM & Avg. & FSH & SWB2 & Avg.\\
\hline\hline
1 & RWTH SMBR BLSTM~\cite{kitza2019cumulative} & - & 6.7 & 14.7 & 10.7 & - & - & -\\
2 & + Affine transform based environment adaptation & - & 6.7 & 13.5 & 10.2 & - & - & -\\
\hline\hline
3 & JHU ESPNet Transformer~\cite{karita2019comparative} & - & 9.0 & 18.1 & 13.6 & - & - & -\\
\hline\hline
4 & Google Listen, Attend and Spell network + SpecAugment~\cite{park2019specaugment} & - & 6.8 & 14.1 & (10.5) & - & - & -\\
\hline\hline
5 & \multirow{3}*{\shortstack{IBM LSTM based Attention encoder-decoder \\+ SpecAugment + weight noise~\cite{tuske2020single}}} & 29M & 7.4 & 14.6 & (11.0) & - & - & -\\
6 & ~ & 75M & 6.8 & 13.4 & (10.1) & - & - & -\\
7 & ~ & 280M & 6.4 & 12.5 & (9.5) & 8.4 & 14.8 & (11.7)\\
\hline\hline
8 & LF-MMI TDNN + BLHUC + Large RNNLM & 19M & 7.6 & 14.4 & 11.0 & 8.9 & 14.3 & 11.7\\
9 & LF-MMI NAS TDNN + BLHUC + Large RNNLM & 13M & 7.1 & 13.3 & 10.2 & 8.5 & 13.9 & 11.3\\
\hline\hline
10 & LF-MMI CNN-TDNN + BLHUC + Large RNNLM & 15.2M & 6.9 & 12.9 & 9.9 & 8.3 & 13.9 & 11.2 \\
11 & LF-MMI NAS CNN-TDNN + BLHUC + Large RNNLM & 10.8M & 6.9 & 13.0 & 9.9 & 8.3 & 13.7 & 11.1\\
\hline\hline
\end{tabular}
\vspace{-2mm}
\end{table*}

\subsubsection{\textbf{TDNN-F Context Offsets and Bottleneck Projection Dimensionality Search}}
The performance of Pipelined Gumbel-Softmax and Pipelined Softmax DARTS methods are further evaluated by searching both context offsets and bottleneck projection dimensionality at each factored TDNN layer. In Tab.~\ref{exp: tdnn_context_offset_bn_dim_results}, their performance are again contrasted with those of the baseline Kaldi recipe system (Sys (1)), the best manually designed system (Sys (2)) and a random search configured system (Sys (3)). The manually designed system (2) is configured using the best-performing context offset and dimensionality settings selected from those manually designed systems in Tab.~\ref{exp: tdnn_context_offset_results} and Tab.~\ref{exp: tdnn_bn_dim_results}. The Random Search system (3) is obtained by selecting the best performing model from 6 randomly sampled models with different context offsets or dimensionality choices during search. 

Systems (4)-(5) in Tab.~\ref{exp: tdnn_context_offset_bn_dim_results} are produced by using the PipeSoftmax or PipeGumbel method to learn the context offsets and bottleneck projection dimensionality settings in a one by one, “two-stage” fashion. This means the architecture search over these two attributes are separately performed in each stage.  In contrast, systems (6)-(7) are produced by using the PipeSoftmax or PipeGumbel method to learn these two attributes jointly in the same super-network during one single round of architecture search. For all systems in Tab.~\ref{exp: tdnn_context_offset_bn_dim_results}, their performance prior to and after further applying additional LHUC~\cite{swietojanski2016learning} speaker adaptation and Kaldi recipe LSTM RNNLM rescoring are also shown.

The details of Kaldi recipe LSTM RNNLM rescoring are as follows: two unidirectional (forward/backward) context based long short-term memory recurrent neural network language models (LSTM LMs)~\cite{sundermeyer2012lstm} were built following the standard Kaldi recipe. The transcripts of both the Switchboard training data and the Fisher English corpora were encoded as 1024 dimensional embedding vectors to train the LSTM LMs. Each LSTM LM consisted of 2 unidirectional LSTM layers with 1024 cells. Projections were used inside the LSTMs to reduce the output dimensions to 512. A context splicing layer with ReLU activation was exploited before and after each LSTM layers. The splicing indices of the three context splicing layers were \{-1, 0\}, \{-3, 0\}, and \{-3, 0\}. Finally, an affine transformation was used to generate the output embedding vectors. For performance evaluation, the two LSTM LMs were employed in turn to rescore the lattices generated by the 4-gram LM.

Several trends can be observed from Tab.~\ref{exp: tdnn_context_offset_bn_dim_results}.
\begin{enumerate}
    \item By searching the context offsets and bottleneck projection dimensionality in a sequential manner, the PipeGumbel Top 1 system (Sys (5)\footnote{NAS selected projection dimensions at each layer:\{80, 80, 80, 50, 80, 50, 80, 50, 80, 80, 100, 120, 120, 160\} and context configurations \{-2,2\}, \{-2,4\}, \{-5,5\}, \{-6,6\}, \{-6,5\}, \{-6,6\}, \{-6,6\}, \{-6,6\}, \{-6,6\}, \{-6,6\}, \{-6,6\}, \{-6,6\}, \{-6,6\}, \{-6,6\}}) produces 0.5\% (\textbf{SWB1} subset of \textbf{Hub5' 00} test set) to 1.2\% (\textbf{SWB5} subset of \textbf{Rt02} test set) absolute WER reductions on three test sets and a relative model size reduction of 31\% over the baseline Kaldi recipe factored TDNN system (Sys (1)).
    \item When compared with the manually designed system (2) in Tab.~\ref{exp: tdnn_context_offset_bn_dim_results}, the PipeGumbel Top 1 system (Sys (5)) achieves comparable performance and reduces the overall system design and training time from 463 to 160 GPU hours. The PipeGumbel Top 1 system (Sys (5)) also outperforms the Random Search system (3) in Tab.~\ref{exp: tdnn_context_offset_bn_dim_results}.
    \item The two stage search over the above two groups of hyper-parameters (PipeGumbel Sys (5)) in practice outperforms performing a joint search over the two (PipeGumbel Sys (7)) by statistically significant absolute WER reduction of 0.3\% on average across three test sets. One possible explanation is that increased modelling confusion and search errors are encountered in the latter case when optimizing both attributes in the same super-network for the PipeGumbel Top 1 system (Sys (7)).
    \item We further hypothesize that by reducing the model size and structural redundancy of TDNN-F systems, the auto-configured NAS systems, for example, using the best performing penalized PipeGumbel system (Sys (5) in Tab.~\ref{exp: tdnn_context_offset_bn_dim_results}), with 12.9M parameters, the uncertainty associated with the resulting TDNN-F model parameters can also be reduced. In order to verify this hypothesis, we apply Bayesian estimation~\cite{mackay1992practical, neal2012bayesian, bishop2006pattern, graves2011practical, pmlr-v37-blundell15,hu2021bayesian} to both the baseline Kaldi recipe configured TDNN-F system and the PipeGumbel NAS configured TDNN-F system (Sys (1) \& Sys (5) in Tab.~\ref{exp: tdnn_context_offset_bn_dim_results}, shown again as Sys (1) \& (2) in Tab.~\ref{exp: tdnn_nas_bayesian}). Efficient variational inference and parameter sampling based training approaches~\cite{hu2021bayesian} were used to estimate the latent parameter posterior distributions in these Bayesian estimated systems. The performance of the corresponding two Bayesian TDNN-F systems are shown as system (3) and (4) in Tab.~\ref{exp: tdnn_nas_bayesian}. It is found that the NAS auto-configured TDNN-F system (Sys (4) in Tab.~\ref{exp: tdnn_nas_bayesian}) has a smaller average parameter latent distribution’s variance (0.025) compared with a larger one (0.033) associated with the baseline system using the baseline recipe configuration (Sys (3) in Tab.~\ref{exp: tdnn_nas_bayesian}). Furthermore, no performance is obtained by applying Bayesian estimation to the NAS auto-configured factored TDNN-F system (Sys (2) vs. (4) in Tab.~\ref{exp: tdnn_nas_bayesian}), in contrast to the significant WER reduction of 0.4\% found on the baseline recipe systems without using NAS (Sys (3) vs. Sys (1)). These results suggest the proposed NAS approaches can effectively minimize the structural redundancy in the TDNN-F systems and reduce their model parameter uncertainty. 
    \item We also report an ablation study on the transferability of NAS configured hyper-parameters across data sets of different sizes, where we contrast the performance of transferring the 300-hour subset set (without speed perturbation) NAS learned TDNN-F context offsets and projection dimensions to the 900-hour full set (with speed perturbation) during system training (Sys (2) in Tab.~\ref{exp: swbd_nas_transferability_result}), against performing both NAS and system training exclusively using the 900-hour full set (Sys (3) in Tab.~\ref{exp: swbd_nas_transferability_result}). The significant WER differences observed between these two systems suggest further research is required to improve the transferability and generalization of NAS methods across different data sets and quantities.
    \item By further incorporating both LHUC speaker adaptation~\cite{swietojanski2016learning} and the Kaldi recipe LSTM RNNLM rescoring, similar performance improvements can still be maintained. Statistically significant absolute WER reductions of 0.5\% (on \textbf{SWB1} subset of \textbf{Hub5’ 00} test set and \textbf{SWB4} subset of \textbf{Rt02} test set) to 1.0\% (on \textbf{CHM} subset of \textbf{Hub5’ 00} test set and \textbf{SWB5} subset of \textbf{Rt02} test set) are obtained by the PipeGumbel Top 1 system (12) over the Kaldi recipe configured baseline system (8). 
\end{enumerate}

In the following experiments, the performance of the best PipeGumbel Top 1 system (Sys (12) in Tab.~\ref{exp: tdnn_context_offset_bn_dim_results}) was further refined by incorporating the more powerful Bayesian LHUC adaptation~\cite{xie2021bayesian} in place of the conventional deterministic LHUC adaptation, and a larger LSTM RNNLM with a doubled number of LSTM cells (2048) and projection dimensionality (1024) compared with the smaller LSTM RNNLM used in Tab.~\ref{exp: tdnn_context_offset_bn_dim_results}. This is shown as system (9) in Tab.~\ref{exp: swbd_nas_newest_result}, together with a comparable PipeGumbel configured CNN TDNN-F system (11)\footnote{NAS selected projection dimensions at each TDNN-F layer: \{120, 100, 80, 80, 80, 80, 100, 100, 120\} and context configurations \{-6,6\}, \{-6,6\}, \{-6,6\}, \{-6,6\}, \{-6,6\}, \{-6,6\}, \{-6,6\}, \{-6,6\}, \{-6,6\}} whose TDNN layer context offsets and projection dimensionality settings are automatically determined. The baseline Kaldi configured TDNN-F and CNN TDNN-F systems are shown as system (8) and (10) respectively. The PipeGumbel NAS configured TDNN-F and CNN TDNN-F systems (Sys (9) \& (11)) are further compared with the state-of-the-art performance obtained using a series of recent hybrid and end-to-end systems reported in the literature (Sys (1)-(7) in Tab.~\ref{exp: swbd_nas_newest_result}). Among these, system (1) and (2) in Tab.~\ref{exp: swbd_nas_newest_result} are the RWTH BLSTM hybrid systems without and with affine transformation for environment adaptation~\cite{kitza2019cumulative}. System (3) is the JHU ESPNet based transformer end-to-end system~\cite{karita2019comparative}. System (4) is the Google Listen, Attend and Spell end-to-end system built with SpecAugment~\cite{park2019specaugment}. System (5)-(7) are the IBM LSTM based attention encoder-decoder end-to-end systems constructed using SpecAugment and weight noise~\cite{tuske2020single}. 

Competitive performance is obtained by the Pipelined Gumbel-Softmax searched TDNN-F and CNN TDNN-F systems (Sys (9) \& (11)) on the \textbf{CHM} subset of \textbf{Hub5’00} test set and \textbf{Rt03S} test sets when compared with the other hybrid and end-to-end systems (Sys (1)-(7) in Tab.~\ref{exp: swbd_nas_newest_result}), while retaining more compact system sizes. In particular, the NAS configured CNN-TDNN-F system (Sys (11) with 10.8M parameters) achieves a state-of-the-art WER of 11.1\% on the \textbf{Rt03S} test set, outperforming the most complex and best performing IBM system (Sys (7) with 280M parameters) by a WER reduction of 0.6\% absolute as well as a 96.1\% relative model size reduction. 

\begin{table*}[t]
\vspace{-2mm}
\caption{Performance (WER\%, number of parameters) of TDNN-F systems configured with both context offsets and projection dimensionality produced by baseline setting, Pipelined Gumbel-Softmax DARTS (PipeGumbel) or Random Search before and after applying LHUC SAT and test time unsupervised adaptation. Random Search selects the best performed model from 6 randomly sampled models (with maximum context offsets $\pm3$). $\eta$ is the penalty factor in Eqn.~(\ref{eq:penalized_darts}). $\{[a,b]\}$:$\{-c, d\}$ denotes context offsets $\{-c,0\}$ to the left and $\{0,d\}$ to the right used from $a$-th layer to $b$-th layer inclusive. The dimensionality index denotes the index of 8 dimensionality choices: \{25,50,80,100,120,160,200,240\}. "Very Low", "Low", "Mid" and "High" denote the group of speakers with different intelligibility. $^\dagger$ denotes a statistically significant difference is obtained over the baseline Kaldi recipe systems (Sys (1) \& (4)).}
\label{exp: uaspeech_tdnn_context_offset_bn_dim_results}
\centering
\begin{tabular}{c|c|c|c|c|c|c|c|c|c|c|c|c}
\hline\hline
\multirow{2}*{Sys} & \multirow{2}*{Method} & \multirow{2}*{SP} & LHUC & \multirow{2}*{$\eta$} & Context offsets \& & \multicolumn{5}{c|}{WER\%} & \multirow{2}*{\#Param} & \multirow{2}*{Time}\\
\cline{7-11}
~ & ~ & ~ & SAT & ~ & Bottleneck Dim Index & Very Low & Low & Mid & High & Avg. & ~ & ~\\
\hline\hline
1 & Baseline (Manual) & \multirow{6}*{\cmark} & \multirow{6}*{\xmark} & \multirow{4}*{-} & \tabincell{c}{\{[1,3]\}:\{-1,1\}; \{4\}:\{0\}; \\ \{[5,7]\}:\{-3,3\} \& 5 5 5 5 5 5 5} & 57.5 & 31.5 & 23.6 & 13.3 & 29.4 & 9.5M & 3h\\
\cline{1-2}\cline{6-13}
2 & \tabincell{c}{Random Search \\(1 best)} & & & &  \tabincell{c}{\{[1,2]\}:\{-3,2\};\{[3]\}:\{-3,1\};\\
\{[4,5]\}:\{-3,3\}; \{[6]\}:\{-2,0\}; \\ \{[7]\}:\{-3,1\} \& 4 3 4 6 7 6 5} & 57.7 & 31.8 & 22.8$^\dagger$ & 14.2 & 29.6 & 9.8M & 17h\\
\cline{1-2}\cline{5-13}
3 & \tabincell{c}{PipeGumbel \\(Top1)} &  & & 0 & \tabincell{c}{\{[1,3]\}:\{0\}; \{[4,7]\}:\{-2,2\} \\ \& 3 4 7 2 3 3 0} & 58.6 & \textbf{30.6}$^\dagger$ & \textbf{21.5}$^\dagger$ & \textbf{11.7}$^\dagger$& \textbf{28.4}$^\dagger$ & 6.4M & 13h\\
\hline\hline
4 & Baseline (Manual) & \multirow{6}*{\cmark} & \multirow{6}*{\cmark} & \multirow{4}*{-} & \tabincell{c}{\{[1,3]\}:\{-1,1\}; \{4\}:\{0\}; \\ \{[5,7]\}:\{-3,3\} \& 5 5 5 5 5 5 5} & 56.1 & 29.6 & 19.5 & 12.8 &27.6 & 9.5M & 7h\\
\cline{1-2}\cline{6-13}
5 & \tabincell{c}{Random Search \\(1 best)} &  &  &  & \tabincell{c}{\{[1,2]\}:\{-3,2\};\{[3]\}:\{-3,1\};\\
\{[4,5]\}:\{-3,3\}; \{[6]\}:\{-2,0\}; \\ \{[7]\}:\{-3,1\} \& 4 3 4 6 7 6 5} & 56.3 & 30.1 & 19.9 & 13.3 & 27.9 & 9.8M & 21h\\
\cline{1-2}\cline{5-13}
6 & \tabincell{c}{PipeGumbel \\(Top1)} & & & 0 & \tabincell{c}{\{[1,3]\}:\{0\}; \{[4,7]\}:\{-2,2\} \\ \& 3 4 7 2 3 3 0} & \textbf{55.8}$^\dagger$ & \textbf{28.0}$^\dagger$ & 19.7 & \textbf{11.2}$^\dagger$ & \textbf{26.6}$^\dagger$ & 6.4M & 17h\\
\hline\hline
\end{tabular}
\vspace{-2mm}
\end{table*}

\subsection{NAS Experiments on Dysarthric UASpeech Task}
~\label{sec: exp_nas_uaspeech}
In order to further evaluate the performance of the proposed NAS techniques, they were applied to automatically configure the above discussed two sets of hyper-parameters of a state-of-the-art dysarthric speech recognition task based on the UASpeech corpus~\cite{kim2008dysarthric}.

\textbf{Task Description:} The UASpeech corpus is the largest publicly available dysarthric speech corpus that is designed as an isolated word recognition task~\cite{kim2008dysarthric}. Approximately 103 hours of speech was recorded from 29 speakers among which 16 are dysarthric speakers while the remaining 13 are healthy control speakers. For speech recognition system development, the entire corpus is further divided into 3 subset blocks per speaker, with each block containing different speech contents based on a mix of common and uncommon words. Among these, the same set of common words contents are used in all three blocks, while the uncommon words in each block are different. The data from Block 1 (B1) and Block 3 (B3) of all the 29 speakers are used as the training set (69.1 hours of audio, 99195 utterances in total), while the data of Block 2 (B2) collected from all the 16 dysarthric speakers (excluding speech from healthy control speakers) serve as the evaluation data set (22.6 hours of audio, 26520 utterances in total). After removing excessive silence at the start and end of speech audio segments~\cite{yu2018development}, a combined total of 30.6 hours of audio data from Block 1 and 3 (99195 utterances) is used as the training set, while 9 hours of speech from Block 2 (26520 utterances) is used for performance evaluation. Similar as Sec.~\ref{sec: exp_nas_swbd}, a GMM-HMM system was trained with 39-dimensional PLP features to generate alignments for the neural network training. The factored TDNN baseline\footnote{The {\tt frames\_per\_eg} variable was set to be 150,110,90,60,30.} featuring speed perturbation\footnote{Following our previous work~\cite{geng2020investigation}, data augmentation featuring speed perturbation of both control and dysarthric speech was conducted, which leads to an augmented training set of 130.1h.} was trained following the Kaldi chain setup, except that i-Vector was not incorporated. Following the configurations specified in~\cite{yu2018development,christensen2013learning}, recognition is performed using a uniform language model with a word grammar network constructed using 255 test set words.

The performance comparison among the baseline TDNN-F system, a random search configured system and the PipeGumbel DARTS auto-configured TDNN-F system, before and after LHUC speaker adaptive training (SAT) and test time unsupervised adaptation were applied, are shown in Tab.~\ref{exp: uaspeech_tdnn_context_offset_bn_dim_results}. The same trend previously observed on the Switchboard task of Sec.~\ref{sec: exp_nas_swbd} can be found again. First, the PipeGumbel auto-configured TDNN-F system (3) consistently outperforms the baseline TDNN-F system (1) and random search configured system (2). For example, statistically significant WER reduction of 1.0\% on average and 33\% relative model size reduction were obtained by PipeGumbel auto-configured TDNN-F system (3) when compared with the baseline system (1). Second, after further incorporating LHUC SAT and test time unsupervised adaptation, similar performance improvements can still be retained. 

\vspace{-2mm}
\section{Conclusion}
This paper presents a range of DARTS based neural architecture search techniques to automatically learn two groups of architecture hyper-parameters that heavily affect the performance and model complexity of state-of-the-art lattice-free MMI trained factored time delay neural network acoustic models: i) the left and right splicing context offsets; and ii) the dimensionality of the bottleneck linear projection at each hidden layer. Parameter sharing among candidate neural architectures was used to facilitate efficient search over a very large number of (up to $7^{28}$) different TDNN-F system configurations.

Experiments conducted on both the benchmark 300-hour Switchboard corpus and the 103-hour UASpeech dysarthric speech recognition task suggest the resulting NAS auto-configured TDNN-F models consistently outperform the baseline systems using manually designed configurations or random search by significant absolute word error rate (WER) reductions up to 1.2\% and model size reduction of 31\% relative. State-of-the-art recognition accuracy performance was obtained on the NIST Hub5’ 00 and Rt03s test sets and compared with those of the most recent hybrid and end-to-end attention and transformer based systems reported in the literature~\cite{kitza2019cumulative,karita2019comparative,park2019specaugment,tuske2020single}. Further analysis using Bayesian learning demonstrates that the proposed set of NAS approaches can effectively minimize the structural redundancy in the TDNN-F systems and reduce their model parameter uncertainty. The generic nature of these techniques also allows their wider application to other speech task domains as well as similar automatic neural architecture configuration problems when developing speech recognition systems using end-to-end approaches. 
\vspace{-3mm}
\section*{Acknowledgment}
This research is supported by Hong Kong Research Grants Council GRF grant No. 14200218, 14200220, 14200021, Theme-based Research Scheme T45-407/19N, Innovation \& Technology Fund grant No. ITS/254/19, PiH/350/20, InP/275/20 and Shun Hing Institute of Advanced Engineering grant No. MMT-p1-19.


\ifCLASSOPTIONcaptionsoff
  \newpage
\fi


\bibliographystyle{IEEEtran}
\bibliography{IEEEexample}

\end{document}